\documentclass{aa} 
\usepackage[varg]{txfonts}
\usepackage{natbib}
\usepackage{graphicx}
\usepackage[colorlinks=true,linkcolor=black,anchorcolor=black,citecolor=blue,filecolor=black,menucolor=black,runcolor=black,urlcolor=black]{hyperref}
\usepackage{subcaption}
\usepackage{float}
\usepackage{url}
\usepackage{xcolor}
\usepackage{arydshln}
\usepackage{placeins}
\makeatletter
\renewcommand*\aa@pageof{, page \thepage{} of \pageref*{LastPage}}
\makeatother

\begin{document}

\title{The Rotation Period Distribution in the Young Open Cluster NGC~6709}
 
\titlerunning{Title Running} 
\authorrunning{Cole-Kodikara et al.}
  
  \author{Elizabeth M. Cole-Kodikara\inst{1}
  \and S. A. Barnes\inst{1,2}
  \and J. Weingrill\inst{1}
  \and T. Granzer\inst{1}
  } 

\institute{
Leibniz-Institute for Astrophysics Potsdam (AIP), An der Sternwarte 16, 14482, Potsdam, Germany
\and
Space Science Institute, Boulder, CO, USA}
 
\abstract{Open clusters serve as a useful tool for calibrating models of the relationship between mass, rotation, and age for stars with an outer convection zone due to the homogeneity of the stars within the cluster.
Cluster to cluster comparisons are essential to determine whether the universality of spin-down relations holds.
NGC\,6709 is chosen as a more distant representative Pleiades-age cluster for which no rotation periods of members have previously been obtained.
This cluster is at a distance of over 1\,kpc and has two red giant members. 
Isochrones place the age of the cluster at around 150\,Myr, or approximately the same age as the Pleiades.
Photometry is obtained over a multi-month observing season at the robotic observatory STELLA.
After basic processing, PSF photometry was derived using \textsc{Daophot II}, and a suite of related software allowed us to create time series of relative magnitude changes for each star.
Four time series analysis methods are then applied to these light curves to obtain rotation periods for members stars. 
We obtain for the first time rotation periods for 45 FGK cluster members of NGC\,6709.
We plot the \textit{Gaia} EDR3 colors of the member stars against their rotation periods and find 
a slow-rotating sequence with increasing rotation periods towards redder stars and a smaller clump of rapid rotators that have not yet joined this sequence.
NGC\,6709 has rotation periods very similar to that of another Pleiades-age open cluster, NGC\,2516.}

\maketitle 


\section{Introduction}\label{sec:intro}

The rotation periods of cool stars, both in open clusters and in the field, have attracted considerable attention in recent years.
Not only does the rotation period provide the basic observable that governs the activity properties of cool stars, but it can also be measured very precisely.
Furthermore, the rotation period distributions evince patterns 
\citep[e.g.,][]{Barnes2003}
that inform us about internal processes in stars or even their ages 
\citep[e.g.,][]{Skumanich1972,Barnes2007}.
Open clusters are particularly valuable as calibrating objects, or more generally because they offer us homogeneous sample of coeval stars of the same composition.
This paper contributes to that effort by presenting a newly determined rotation period distribution for a hitherto unstudied young open cluster NGC\,6709. 

The Sun is not unique as a 
star, and so the physics of the 
solar-wind and angular momentum of the Sun in studies by those 
such as 
\cite{Parker1958} and \cite{Weber1967}
can be applied to stars.
A star after formation should spin down as angular 
momentum is lost from the interactions of the stellar winds
with magnetic fields \citep[e.g.,][]{Mestel1968a}. 
This tendency for older stars to spin slower takes the form 
of $v \sin i \propto t^{-1/2}$ for solar mass stars, 
or Skumanich's relationship \citep{Skumanich1972}. 
This relationship has been generalized to one where the rotation period depends on both the mass (or equivalently color) and age of a star.
If such a relationship actually exists, then it can be inverted to provide the age of a star from its measured rotation period and color, a procedure that is called gyrochronology \citep{Barnes2003}.

Early measurements of stellar rotation predating large-format 
CCDs used $v \sin i$ measurements from spectroscopy \citep[e.g.,][]{Gaige1993}.
Modern observing techniques allow more precise and unambiguous 
measurements of the rotation period via photometric changes.
Thus, the most efficient way of observing the rotation rates of stars is through changes in brightness in a time series of observations of an object.
\cite{Kron1947} initially observed periodic changes in light curves due to large, cool star spots.
FGKM stars have an outer convection zone, and the suppression of convection from the magnetic field causes a cool spot to form, analogous to the sunspots. 
As these spots rotate across the line of sight, a dip 
in the stellar brightness may be observed \citep[e.g.,][]{Strassmeier2009}. 
Since starspots are thought to last at least the order of a 
few months, the stellar rotation can then be extracted 
from a time series of observations of a star's brightness.

Open clusters provide a rich laboratory for observing stellar rotation.
Cluster members can be considered largely homogeneous in 
chemical composition and age \citep[e.g.,][]{DeSilva2006},
and so observing the rotation rates of the 
FGK stars (and M-dwarfs where possible) 
can provide information about the dependence of rotation periods on mass.
Furthermore, the age of the open cluster can be determined by 
other independent methods 
such as isochrones \citep[e.g.,][]{Sandage1969} and measurements of the chromospheric flux
\citep[e.g.,][]{Mamajek2008}. 
Open clusters can therefore serve to calibrate the age-mass-rotation relationship 
essential to gyrochronology.
This method of using individual open clusters as snapshots 
of stellar rotation at a single age only holds if 
all clusters of a similar age and metallicity show similar distributions 
in rotation periods.
Cluster to cluster comparisons that become essential to this theory.

One benchmark candidate for young open clusters around the age 125\,Myr 
is the well-studied Pleiades at a distance of 133\,pc 
\citep[e.g.,][]{vanLeeuwen1987,Hartman2010,Rebull2016},.
This cluster, along with the Hyades \citep[e.g.,][]{Radick1987,Delorme2011,Douglas2019},
was used to derive the Skumanich's relationship for the evolution of rotation for 
solar-type stars.  
\cite{Fritzewski2020} (hereafter F20) compare the rotation rates of members of 
four other Pleiades-age clusters and find that all measured ones behave similarly.

NGC\,6709 
was chosen as a significantly more distant representative of 
Pleiades-age clusters with no previous observations of rotation 
rates of member stars.
Several membership studies are available the in literature 
relying on either spectroscopic or proper motion membership 
determination methods
\citep{Hakkila1983,Tadross2001,Dias2001,Kharchenko2004,Putte2010,
Kharchenko2013,Dias2014,Sampedro2017,CantatGaudin2018}.
It has two red giant members 
\citep{Lindoff1968,Sears1997,Ahumada2007,Mermilliod2008}, 
possible Be stars \citep{Schild1976,Subramaniam1999},
and possible blue stragglers \citep{Ahumada2007}.
One of the red giants is a spectroscopic binary 
\citep{Mermilliod2007} and the other is a possible member
of a quadruple star system \citep{Smiljanic2018}.

The above-1000\,pc distance and consequent faintness of the cluster has led some disparity 
in adopted parameters due to the difficulty in observing 
the cluster, although with the advent of \textit{Gaia},
these values have been more in agreement in recent publications.
Values for the $E_{B-V}$ color excess  range from 0.28--0.35
\citep{Johnson1961,Becker1963,Burki1975,Mermilliod1981,
Subramaniam1999,Tadross2001,Lata2002,Ahumada2007,Piskunov2007,
Kharchenko2013,Joshi2016,Sampedro2017}.
Distance estimates range from 930--1253\,pc 
\citep{Johnson1961,Hoag1965,Lindoff1968,Burki1975,Mermilliod1981,
Hakkila1983,Barkhatova1987,Lynga1987,Leisawitz1988,Subramaniam1999,
Dias2001,Tadross2001,Loktin2003,Kharchenko2005,CantatGaudin2018}.

Age estimates based on various methods of isochrone fitting 
in literature have ranged from 65\,Myr to 320\,Myr.
Earlier studies of NGC\,6709 tend to obtain ages
of less than 80\,Myr 
\citep{Lindoff1968,Burki1975,Mermilliod1981,Barkhatova1989}
or older than 200\,Myr \citep{Barbaro1969,Sagar1999,Subramaniam1999},
possibly due to instrument limitations at the time and the distance 
to NGC\,6709.  
More recent studies place the age around 100--190\,Myr 
\citep{Tadross2001,Lata2002,Chen2003,Kharchenko2005,Ahumada2007,
Wu2009,Putte2010,Smiljanic2018,CantatGaudin2020,Bossini2019}. 

Radial velocity measurements have only been obtained 
for a few member stars, including the red giant members.
Hence, there is some degree of uncertainty, and published 
values range from $-7$ to $-13$\,km$\cdot$s$^{-1}$
\citep{Lynga1987,Kharchenko2005,Putte2010,Kharchenko2013,Conrad2017,Soubiran2018}.
Metallicity is not well-measured for this cluster, but
spectroscopy from UVES members from \textit{Gaia}-ESO data 
have the cluster near solar values \citep{Magrini2021}.

This work is part of the 
STELLA Open Cluster Survey (SOCS). 
The survey observes multiple open clusters 
over periods of several months to obtain rotation periods.
The objective is to establish the similarities 
between open clusters of similar ages as well as 
track the evolution of rotation periods over time 
with clusters of different ages \citep[e.g.,][]{Feugner2011,Barnes2015}. 

\section{Methods} \label{sec:methods}

In outline, to derive rotation periods for stars in NGC\,6709,
we observe the field over several months, 
a comparatively long baseline that is permitted by our robotic observing, 
determine which stars in the field belong 
to the cluster, and obtain their corresponding colors from literature. 
We then construct light curves for each star from the collection of frames.
We then perform time series analyses to obtain rotation periods 
and error estimates for each member star.

\subsection{Data Aquisition and Reduction}\label{sec:data}

\begin{table}
\centering
\caption{Observations and Exposure times}
\begin{tabular}{lccc}
\hline \hline
Exposure &  Start Date   &  End Date & Number of \\
time &    &   & observations \\
$[$s$]$ &  yyyy-mm-dd   &   yyyy-mm-dd & \\
\hline
 24    & 2016-04-16 & 2016-09-05 & 228 \\
120     & 2016-04-20 & 2016-08-04 & 167 \\
300     & 2015-05-26 & 2015-11-07 & 82 \\
\hline
\end{tabular}
\label{tab:observations}
\end{table}

The observations were obtained using the 1.2\,m telescope STELLA-I at the Iza\~na Observatory in Tenerife, Spain from May 2015 until September 2016.
This is part of the STELLA robotic observatory which also includes
the 1.2\,m spectroscopic telescope STELLA-II.
The observatory operates robotically each night via in-house custom-built software
\citep{Granzer2004},
provided conditions are suitable for observing targets determined by 
a program with user-defined priorities. 
For more information about STELLA operation, please see
\cite{Strassmeier2004}, 
\cite{Strassmeier2010}, and \cite{Weber2016}. 

The CCD resolution for the wide-field imager WiFSIP V2.0 
is $4066 \times 4064$ pixels, or $22\arcmin \times 22\arcmin$. 
A single field was used to observe NGC\,6709 due to its relatively 
small angular size.
This field contains over two-thirds of the member stars when centered on the cluster. 
Observations were acquired in the Johnson $V$ band with exposure times of 24\,s, 120\,s, and 300\,s to  
retain sensitivity to as wide a range of bright (bluer) and faint (redder) stars as was practicable.

Table \ref{tab:observations} summarizes the dates and total number of successful observations available
for each exposure time,
after discarding any unusable frames due to bad seeing, 
poor weather conditions, lunar proximity, and tracking errors.
Significant numbers of such frames are a typical consequence of robotic observations.
With the 24\,s exposures, visits to the target happened up to 
ten times per night, although typically usable frames on 
any successful night were in the range of three to seven.
For the longer exposures of 120\,s, this dropped to a maximum of 
seven times per night, with typical values below six.
And for the longest exposure time, the range was up to four times
per night for the target.
The maximum time baseline of observations for each exposure group is under six months to 
observe multiple theoretical variability cycles.
This dataset exceeds our minimum requirement of 50 usable frames 
for each exposure time and the minimum acceptable time baseline of 60 days 
to preserve our sensitivity to the longer rotation periods.

\subsubsection{Membership}\label{sec:membership}

\begin{table}
\centering
\caption{Properties of open cluster NGC 6709}
\begin{tabular}{rlc}
\hline \hline
Parameter &Value & Reference \\
\hline
Right Ascension & RA = 282.836\,deg & 1 \\
Declination & Dec = +10.334\,deg & 1 \\
Proper Motion & $\mu_{\rm{RA}}$  = 1.41\,mas/yr & 1\\
              & $\mu_{\rm{Dec}}$ = -3.53\,mas/yr&  1 \\
Parallax & $\pi$ = 0.911\,mas & 1 \\
Radial Velocity & $v_{\rm{rad}}$ = -10.2\,km/s& 2 \\ 
Age & $T = 150$\,Myr & 3\\
\hline
\end{tabular}
\tablefoot{(1) \cite{CantatGaudin2018}, (2) \cite{Soubiran2018}, 
(3) \cite{Sampedro2017}}
\label{tab:properties}
\end{table}

We adopt the membership study of \cite{CantatGaudin2018} (hereafter CG18)
to define member stars as all stars in the study with a 
membership probability of $50\%$ or greater.
This study makes use of the recent high-precision \textit{Gaia} observations
and includes many dimmer members that previous studies were unable to 
observe likely due to instrument limitations
\cite[e.g.,][]{Hakkila1983,Kharchenko2013,Dias2014}.
According to CG18, the mean proper motion of NGC\,6709 is 
$\mu_\mathrm{RA} = 1.414$\,mas\,yr$^{-1}$ and
$\mu_\mathrm{Dec} = -3.532$\,mas\,yr$^{-1}$, 
and the distance is 1064\,pc.  
The properties we adopt for NGC\,6709 and references are listed in 
Table \ref{tab:properties}.
The study by CG18 with our membership criteria 
provides 299 open cluster members total.

\begin{figure}
    \centering
    \includegraphics[width=1\columnwidth]{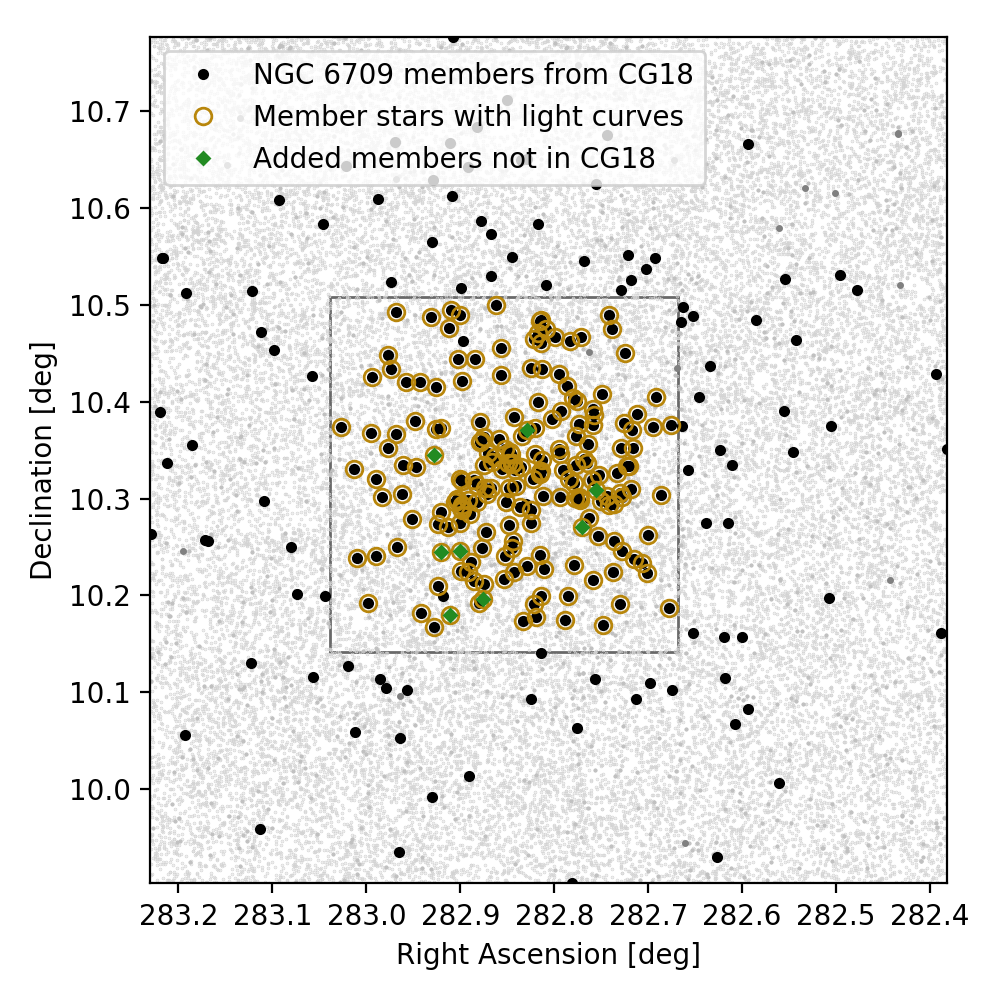}
    \caption{NGC 6709 members. The field of observation is centered 
    on the dense center of the cluster and encompasses approximately two-thirds
    of the members.
    Stars used in this study are encircled, and our field of view 
    is indicated by the square. Additional stars used as members in this study are shown as green diamonds.
    }
    \label{fig:map}
\end{figure}

\begin{table*}
\centering
\caption{Additional cluster members}
\begin{tabular}{lcccccc}
\hline \hline
Designation & RA$^{(1)}$ & Dec$^{(1)}$ & $\mu_{\rm{RA}}$  & $\mu_{\rm{Dec}}$ & $\pi$ & Prob$^{(2)}$ \\
\textit{Gaia} EDR3 & [deg] & [deg] & [mas/yr] & [mas/yr] & [mas] &  \\
\hline
4311548412384207488 & 282.911 & 10.180 & $1.69\pm0.08$ & $-3.53\pm0.08$ & $0.87\pm0.08$ & 0.7\\
4311549138251202560 & 282.875 & 10.196 & $1.36\pm0.19$ & $-3.25\pm0.18$ & $0.85\pm0.18$ & 0.5\\
4311549962915559168 & 282.900 & 10.246 & $1.24\pm0.09$ & $-3.54\pm0.10$ & $0.95\pm0.08$ & 0.7\\
4311549997275268352 & 282.921 & 10.245 & $1.50\pm0.08$ & $-3.40\pm0.07$ & $0.83\pm0.07$ & 0.8\\
4311553708127246848 & 282.771 & 10.270 & $1.49\pm0.15$ & $-3.53\pm0.16$ & $0.94\pm0.17$ & 0.8\\
4311559824161184128 & 282.928 & 10.345 & $1.40\pm0.15$ & $-3.57\pm0.17$ & $0.89\pm0.14$ & 0.8\\
4311600918410547712 & 282.755 & 10.309 & $1.46\pm0.03$ & $-3.73\pm0.03$ & $0.89\pm0.04$ & 0.6\\
4311607373703421056 & 282.829 & 10.371 & $1.67\pm0.03$ & $-3.40\pm0.03$ & $1.04\pm0.03$ & 0.6\\
\hline
\end{tabular}
\tablefoot{(1) ICRS coordinates. (2) Membership probabilities calculated using \cite{Sagar1987}.}
\label{tab:addmembers}
\end{table*}

Figure \ref{fig:map} shows the location of all stars with
membership probability $\ge 0.5$ and the stars used in this study.
Our field of observation covers the denser cluster center, and so 
we have observations for 210 of the cluster members. 
CG18 was limited to stars  
with $m_G < 18$ due to the increase in errors at 
higher apparent magnitudes.
This effectively limits the determination of membership for stars 
of NGC\,6709 to K-stars or bluer due to the distance and 
reddening of the cluster. 
We search for possible additional members not included in CG18
using the \textit{Gaia} EDR3
proper motion data and membership equation of \cite{Sagar1987}.
These additional stars are listed in Table \ref{tab:addmembers}
and are the circled diamonds in Figure \ref{fig:map}.
Radial velocity measurements for cluster members are sparse,
with measurements mainly obtained from the red giants and 
a few other bright members
\citep[e.g.,][]
{Subramaniam1999,Mermilliod2008,Kharchenko2013,Soubiran2018} 
and so membership through radial velocity measurements is not possible with 
existing surveys.

\subsection{Catalogs}\label{sec:outsidedata}

Supporting data for observations in standard passbands 
are taken from the literature by cross-referencing the coordinates of stars in our field with
widely used catalogs.
The coordinates were obtained by converting pixel coordinates 
to sky coordinates using \textsc{WCSTools}
\citep{WCSTools}
and then matched to catalogs with  
\textsc{Astropy} \citep{ASTROPY}.
We use the UCAC4 catalog 
for measurements in the  $B$ and $V$ passbands \citep{Zacharias2013}
and the \textit{Gaia} EDR3 catalog
for $G$, $G_{BP}$, and $G_{RP}$ 
s\citep{GAIA2016,GAIA2020}.
We find some improvement in the \textit{Gaia} colors 
with the EDR3 catalog over the DR2 values given in CG18.

\begin{figure}
\includegraphics[width=\columnwidth]{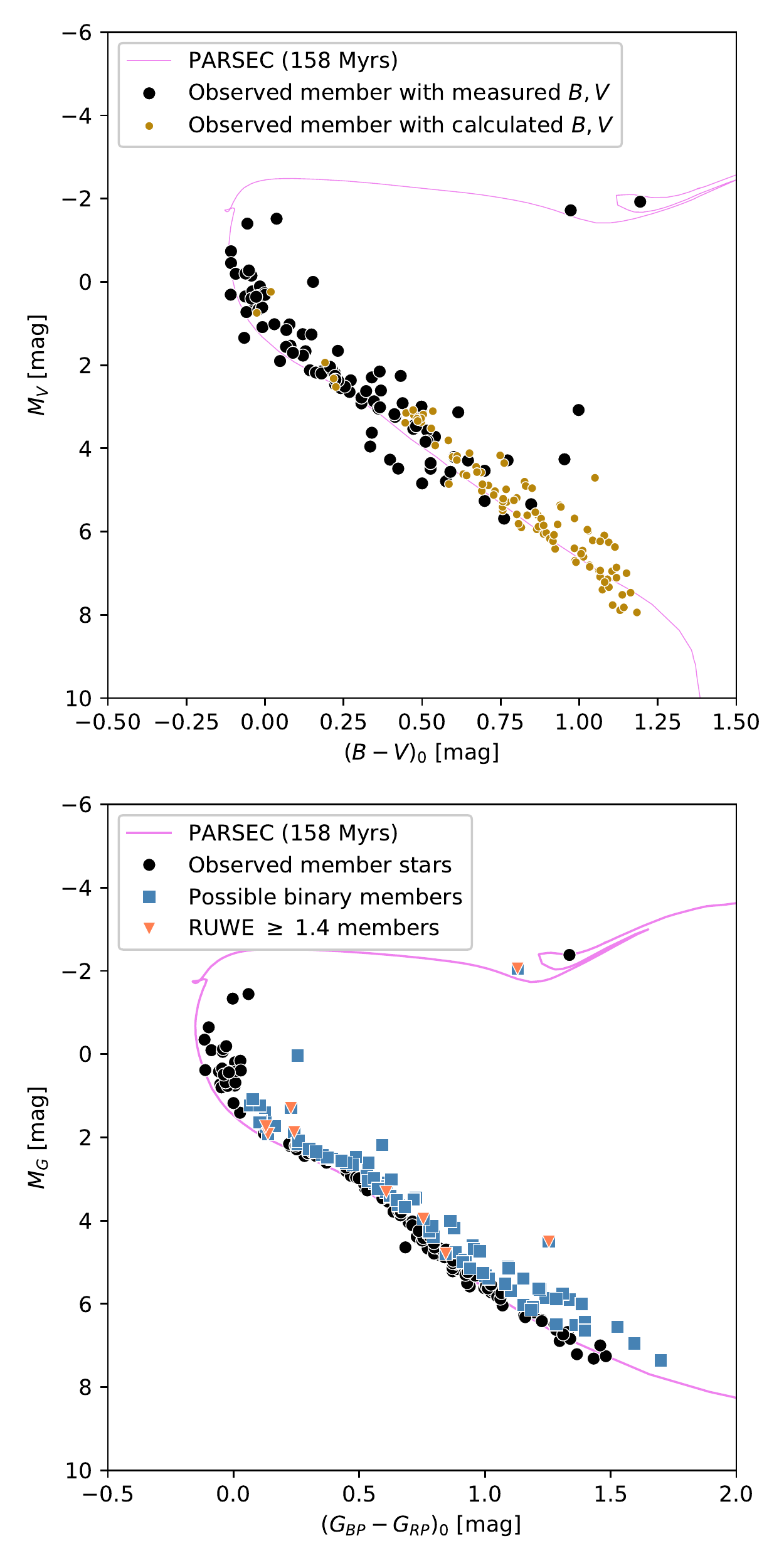}
\caption{Color magnitude diagrams for NGC 6709 
in $B-V$ (above), and Gaia color (below). 
PARSEC isochrones for $\log$Age = 8.2 are also displayed. 
The line-of-sight extinction is calculated for each star using individual stellar 
parallaxes as measured by \textit{Gaia} instead of the mean cluster 
distance. 
Calculated $(B-V)_0$ from \textit{Gaia} colors are shown in gold (above).
Photometric binaries are indicated in blue and RUWE-value binaries 
are indicated in red (below).
}\label{fig:CMD}
\end{figure}

We primarily utilize the \textit{Gaia} EDR3 colors for this work as it is the most complete catalog to date with photometry for almost all of the stars in our field. 
Observations in the $B$ and $V$ passbands only exist for the brighter stars and hence are available for only a fraction of our cluster members.
In order to facilitate a comparison with studies performed in $B-V$ colors, 
we use the equations in Appendix A of \cite{Gruner2020} 
to calculate values for $B-V$ from $G_{BP}-G_{RP}$ when no such values can be found in the 
UCAC4 catalog.
The conversions between the two are piece-wise empirical functions fit to, among others, the Pleiades and so should be reasonable for NGC\,6709.

\subsubsection{Open Cluster Parameters of NGC 6709}\label{sec:ocparameters}

\begin{table}
\centering
\caption{Adopted Parameters used for NGC 6709}
\begin{tabular}{rl}
\hline \hline
Parameter &Value\\
\hline
Metallicity & Z = 0.013 \\
Reddening & $E(G_{BP}-G_{RP})$ = 0.39\,mag \\
 & $E(B-V)$ = 0.29\,mag \\
Extinction & $A_G$ = 0.78\,mag \\
 & $A_V$ = 0.90\,mag\\ 
\hline
\end{tabular}
\label{tab:adoptedparameters}
\end{table}

Values for reddening are selected based on literature 
values and adjusted by fitting isochrones.
Literature values fall in the relative range from 0.28 \citep{Kharchenko2013} to 0.35
\citep{Subramaniam1999} for $E(B-V)$.
We use the relations from \cite{Gruner2020} of:
\begin{equation}
    E(G_{BP}-G_{RP}) = 1.339 \cdot E(B-V),
\end{equation}
\begin{equation}
    A_G = 2.0 \cdot E(G_{BP}-G_{RP}), 
\end{equation}
and 
\begin{equation}
    A_V = 3.1 \cdot E(B-V).
\end{equation}
\textit{Gaia} EDR3 measured parallaxes for each star 
are used for the 
distance modulus calculation, rather than a single value 
for the entire cluster. 

We use a fit to solar-metallicity \textsc{PARSEC} isochrones for 158\,Myr to 
determine a good reddening value for this work \citep{Bressan2012}. 
We select solar metallicity based on the spectroscopic analysis
of \cite{Smiljanic2018} 
for the red giant member BD+10 3697 (\textit{Gaia} EDR3 4311559274405377280), 
where [Fe/H] $= -0.01$.
This isochrone also provides a good fit for the two 
red giants in the membership study of CG18.
The CMDs corrected with the values in Table \ref{tab:adoptedparameters}
are shown in Figure \ref{fig:CMD} for $(B-V)_0$ 
versus $M_V$ and $(G_{BP}-G_{RP})_0$ versus $M_G$.
Where measurements are lacking for $M_B$ and $M_V$, 
we instead use the conversion of \cite{Gruner2020}
to convert the \textit{Gaia} colors to Johnson colors
and fill in the lower 
right of the main sequence in the upper panel as well as a few
brighter members which also lack observations in both $B$ and $V$.
Also shown is the \textsc{PARSEC} isochrone.
Binaries in the lower panel are either determined by their 
location above the main sequence or by the corresponding \textit{Gaia} 
re-normalized unit weight error measurements 
where $RUWE < 1.4$ indicates a good astrometric 
solution and hence single star, and values above this
indicate a possible multistar system \citep{GAIA2016}.
We indicate which type of binary (photometric or RUWE) 
in Figure \ref{fig:CMD}, bottom.

\subsection{Construction of Light Curves}\label{sec:constructlc}

The light curves for individual stars are constructed using  
\textsc{Daophot II} \citep{Stetson1987,Stetson1990}.
First, one master frame for each exposure time was 
selected from all the usable observations.
We selected frames from the following dates: 
2016-07-09 for 24\,s, 
2016-05-31 for 120\,s, 
and 2015-07-17 for 300\,s.
These frames were selected based on low levels of light contamination,
the goodness of the instrument focus, the lack of artifacts, and 
the low ellipticity of the stars. 
In order to use point-spread-function (PSF) photometry,
suitable stars had to be selected from all possibilities found 
by \textsc{Daophot II}.
Approximately 300 sources were selected for PSF photometry 
from the suggested PSF stars found by the program, and 
individually checked for 
counts comfortably between the background 
and saturation level and a minimum distance from other light sources 
of at least $15\arcsec$.
Furthermore, a cross section examination of pixel values across these stars 
should have a single peak and the star circular.
As a final step, any stars located near bad pixels were discarded.

\begin{figure}
    \centering
    \includegraphics[width=1\columnwidth]{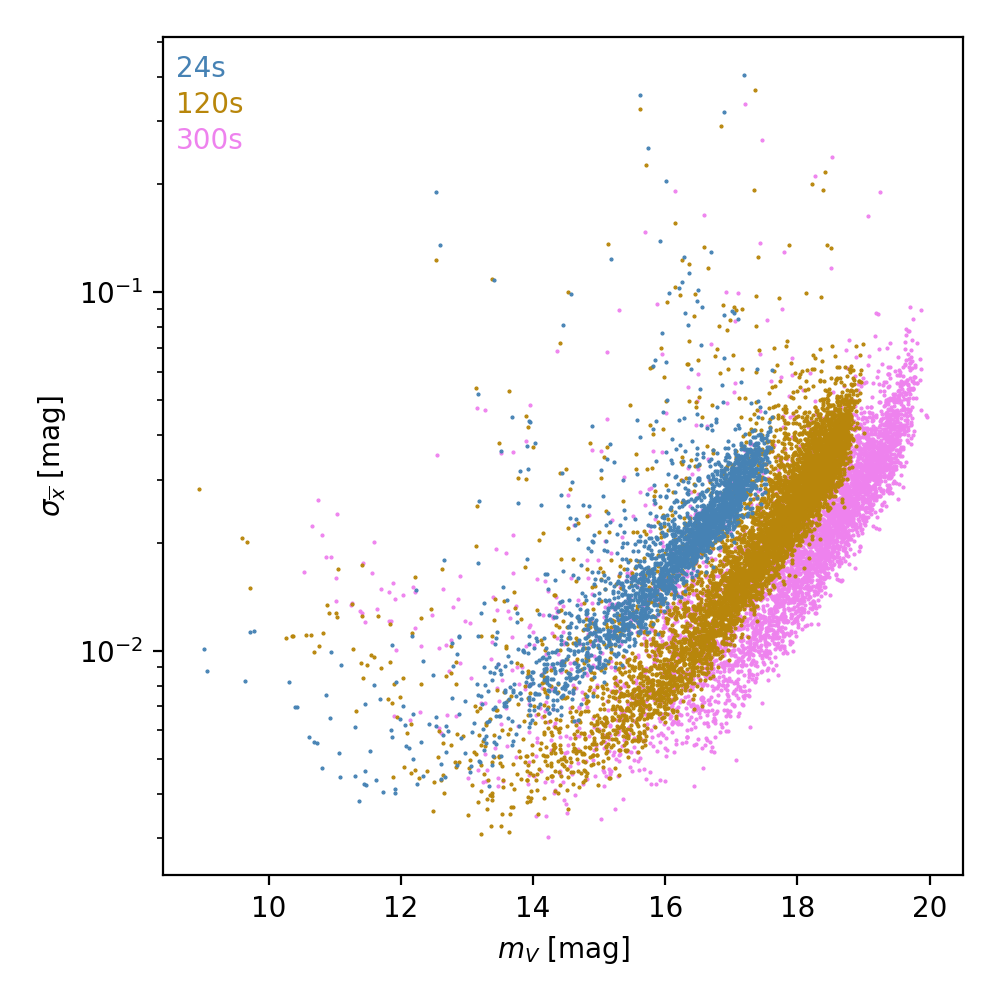}
    \caption{The mean magnitude versus standard error
    for all stars with light curves. 
    }
    \label{fig:variance}
\end{figure}

Once the stars for PSF photometry are selected, 
\textsc{AllStar} is used to find instrumental magnitudes for 
all sources in each 
frame using 
the stars assigned for the PSF photometry as a reference point. 
These magnitudes from each of these frames are then assembled as 
light curves using 
\textsc{DaoMatch} and \textsc{DaoMaster}.
\textsc{DaoMatch} first pinpoints each source possible in every 
frame by taking the 30 brightest stars in each separate frame,
matching triangles with the master frame, and
deriving a coordinate transformation for use from frame to frame.
\textsc{DaoMaster} then refines those transformations 
and applies them to all sources and assigns  
each source its own identifier across all frames.
Each source is then outputted with its pixel location on the master frame 
and measured magnitudes from each
frame, forming a time series for each star. 

Figure \ref{fig:variance} shows the mean apparent magnitude
found by \textsc{Daophot II} 
versus the standard error $\sigma_{\overline{x}}$.
Magnitude differences are detectable down to the millimag regime. 
The shift in the apparent magnitude access for each exposure set 
is the average of the difference between the $m_V$ determined by 
\textsc{Daophot II} and the apparent magnitude in the $V$-band
for stars in the \textit{UCAC4} catalog.
It can be seen that the fainter stars overall exhibit more uncertainty,
and that the field contains some highly variable stars.
The final step in producing the light curves was to remove outliers 
using the sigma clipping function in \textsc{SciPy} \citep{SCIPY,Virtanen2020}. 

\subsubsection{Time Series Analysis} \label{sec:tsa}

To analyze the light curves, four methods were selected with the criteria 
of applicability to randomly spaced observations. 
The Phase Dispersion Minimization method (hereafter PDM) from 
\cite{Stellingwerf1978} tests for periods that minimize the dispersion 
when the data are phased to the test periods.
No a priori knowledge of the shape of the light curve is required for this 
method and so possible spot structure should not affect the results.
The String Length method (hereafter SL) of \cite{Dworetsky1983} phases the 
data with a test period and calculates a total ``string length'' between
the data points, seeking periods that minimize this string length.
Similar to the PDM, this method utilizes phase folding and seeks a minimum in
the periodogram.
This method is chosen as an alternative method to the PDM 
to examine the dispersion present in a phase-folded light curve.

The Generalized Lomb-Scargle method (or GLS) of \cite{Zechmeister2009}
is a frequency analysis method that accounts for errors and offsets 
of the dataset.
The GLS method can handle unevenly sampled data and fits the light curves 
to a combination of sine and cosine waves.
We additionally use the CLEAN method of \cite{Roberts1987} which also samples
frequencies but removes 
artifacts from unevenly sampled data such as the one-day alias.
This serves as a complementary method to GLS in that 
it may amplify the true signal when more than one period is present in the 
data or there is a contribution of noise. 
Periods obtained via these methods appear as maxima in the periodograms.

\subsubsection{Period Selection} \label{sec:periodselection}

After obtaining periodograms with the four methods, we select 
a period based on the following criteria:
\begin{enumerate}
    \item There is agreement within 0.1 days between a primary dip in 
    the PDM method and primary peak in the GLS method. There is similar 
    evidence of minima/maxima in the periodograms of
    the SL and CLEAN methods.
    \item The magnitude variations are higher than the noise from the 
    constant stars of similar magnitudes.
    \item The phase-folded data have a single maximum and minimum to 
    avoid selecting a higher harmonic of the period.
    \item The fit is valid across the most of the time series and there is a suitable density of observations.
\end{enumerate}
Each criterion is given one point, and only periods that have three or four points total 
for at least one exposure time are selected as candidate periods for the stars. 

The light curve and the selected period are visually 
inspected for each of the three exposure times independently.
This system of determining periods 
tends to err on the side of caution in assigning a rotation period
since each individual method does not always successfully recover 
the true rotation period and depends on factors such as the signal-to-noise 
ratio, object brightness, and sampling \citep[e.g.,][]{Graham2013}.
186 out of the 204 cluster member stars were found in more than one exposure batch,
and so the selected periods at different exposure times for the same star are
compared.
In most cases, the discrepancy is less than 0.1 day, and the few exceptions are 
harmonics or subharmonics.

From the cases where a star has more than one possible period due to the 
multiple exposure times, we select as our period whichever  
fulfills more criteria, has the clearest peaks or dips in the periodograms,
and has the most observations, in that order. 
In some cases, a period is not recovered at a different exposure time
either because the 
exposure time was too short to capture the magnitude changes and variability,
or the exposure time was too long and reached saturation. 

\subsubsection{Uncertainties in the Period} \label{sec:errors}

Error bars for periods are obtained from Gaussian fits to the corresponding period 
minimum in the PDM. 
Fits are made for PDMs with minima of 0.6, although exceptions are made 
for for those with a minimum between 0.6 and 0.8 that show a strong 
single peak in the GLS periodogram.
We use the standard deviation $\sigma$ of the Gaussian distribution
as an indication of the uncertainties in the periods so that wide peaks
correspond to a larger uncertainty than narrow peaks.
We use the only the error associated with the exposure the final period was selected from.
Because this is selected via Gaussian distributions fit to the PDM periodogram, 
this $\sigma$ is accordingly only associated with the PDM method. 
The SL method does not provide clear minima in most cases, and the 
GLS and CLEAN methods use a grid in frequency space and thus skew the 
errors when translated to the period space. 

\section{Results} \label{sec:results} 

\begin{table*}
\centering
\caption{Results for NGC\,6709 Member Stars with $P_\mathrm{rot}$}
\begin{tabular}{lcccccccc}
\hline \hline
Designation & RA$^{(1)}$ & Dec$^{(1)}$ & $(G_{BP}-G_{RP})_0$ & $(B-V)_0$ & 
$(B-V)_\mathrm{calc}\,^{(2)}$ & 
Prob$_\mathrm{CG18}\,^{(3)}$ &  $P_\mathrm{rot}$ & $\sigma$ \\
\textit{Gaia} EDR3 & [deg] & [deg] & [mag] & [mag] & [mag] &   &  [d] & [d] \\
\hline
4311605728774145024 & 282.692 & +10.405 & 0.872 &  	& 0.692 & 1.0 & 5.59 & 0.22 \\ 
4311599166063884288* & 282.701 & +10.263 & 1.361 &  	& 1.134 & 0.8 & 7.42 & 0.37 \\ 
4311552127579306624 & 282.706 & +10.234 & 0.891 & 0.953  &  & 0.9 & 5.46 & 0.21 \\ 
4311602017922290176 & 282.716 & +10.352 & 0.772 & 0.576  &  & 0.6 & 4.22 & 0.08 \\ 
4311552161939033984 & 282.728 & +10.246 & 0.819 & 0.602  &  & 1.0 & 4.69 & 0.19 \\ 
4311601983562535936 & 282.729 & +10.353 & 0.798 &  	& 0.626 & 1.0 & 4.36 & 0.18 \\ 
\hline
4311551371664962048 & 282.730 & +10.191 & 1.276 &  	& 1.061 & 0.9 & 5.54 & 0.34 \\ 
4311600780971586048 & 282.741 & +10.294 & 0.871 &  	& 0.692 & 1.0 & 5.63 & 0.17 \\ 
4311600746611842816 & 282.751 & +10.298 & 0.745 &  	& 0.579 & 0.9 & 2.82 & 0.11 \\ 
4311601120223606400* & 282.752 & +10.324 & 1.242 &  	& 1.030 & 1.0 & 3.81 & 0.05 \\ 
4311552230658504064 & 282.753 & +10.261 & 1.062 &  	& 0.866 & 0.9 & 2.26 & 0.02 \\ 
4311602567678117248 & 282.759 & +10.393 & 1.070 &  	& 0.874 & 0.9 & 7.86 & 0.31 \\
\hline
4311601536885870976 & 282.765 & +10.337 & 0.696 &  	& 0.538 & 1.0 & 0.71 & 0.10 \\ 
4311611879167920896 & 282.771 & +10.466 & 0.886 &  	& 0.705 & 0.9 & 5.74 & 0.26 \\ 
4311553914285720448* & 282.774 & +10.301 & 0.595 &  	& 0.452 & 0.9 & 2.88 & 0.06 \\ 
4311608477553696768 & 282.775 & +10.401 & 1.015 &  	& 0.823 & 0.9 & 9.01 & 0.22 \\ 
4311601330727819648 & 282.776 & +10.334 & 0.872 &  	& 0.692 & 1.0 & 5.66 & 0.18 \\ 
4311601743044328704 & 282.777 & +10.365 & 0.840 &  	& 0.663 & 1.0 & 1.01 & 0.10 \\ 
\hline
4311553914285717632 & 282.778 & +10.301 & 0.821 &  	& 0.646 & 1.0 & 4.46 & 0.12 \\ 
4311601262008339968 & 282.785 & +10.321 & 0.800 &  	& 0.628 & 0.9 & 4.63 & 0.11 \\ 
4311550993707780224 & 282.786 & +10.199 & 0.845 & 0.699  &  & 0.8 & 4.78 & 0.13 \\ 
4311608511913447680* & 282.786 & +10.416 & 0.922 &  	& 0.738 & 0.6 &12.46 & 0.92 \\ 
4311609233467993216 & 282.814 & +10.461 & 1.034 &  	& 0.841 & 0.9 & 6.97 & 0.33 \\ 
4311607687279658496 & 282.817 & +10.400 & 0.873 &  	& 0.693 & 0.9 & 5.64 & 0.17 \\
\hline
4311609267827745792 & 282.817 & +10.470 & 0.750 & 0.500  &  & 0.9 & 2.65 & 0.04 \\ 
4311552436816786304 & 282.829 & +10.230 & 0.961 &  	& 0.774 & 0.9 & 4.50 & 0.16 \\
4311547798252006272 & 282.834 & +10.174 & 0.447 & 0.307 	&  & 0.9 & 1.79 & 0.02 \\
4311607309322456704* & 282.835 & +10.365 & 0.912 &  	& 0.729 & 0.9 & 5.60 & 0.15 \\ 
4311554189163566720 & 282.840 & +10.313 & 0.923 &  	& 0.738 & 0.9 & 5.30 & 0.17 \\ 
4311552780414370944 & 282.845 & +10.249 & 0.756 & 0.772  &  & 0.9 & 2.86 & 0.09 \\ 
\hline
4311560305197607296* & 282.863 & +10.343 & 1.090 &  	& 0.892 & 0.9 & 2.28 & 0.02 \\ 
4311553398889528320* & 282.872 & +10.305 & 1.002 &  	& 0.811 & 1.0 & 6.12 & 0.10 \\ 
4311608301417284736 & 282.885 & +10.444 & 0.932 & 0.761  &  & 1.0 & 5.97 & 0.25 \\ 
4311553330139990784 & 282.892 & +10.298 & 1.049 &  	& 0.854 & 0.7 & 6.95 & 0.29 \\ 
4311607996517234048 & 282.898 & +10.421 & 1.061 &  	& 0.866 & 0.9 & 6.97 & 0.27 \\ 
4311549859836309248 & 282.899 & +10.225 & 1.156 &  	& 0.952 & 1.0 & 3.09 & 0.05 \\
\hline
4311553124011562368* & 282.899 & +10.290 & 1.308 &  	& 1.089 & 0.6 & 7.36 & 0.21 \\ 
4311610745296412800* & 282.901 & +10.489 & 1.153 &  	& 0.950 & 1.0 & 7.95 & 0.42 \\ 
4311559343130336128* & 282.911 & +10.324 & 0.925 &  	& 0.740 & 0.8 & 5.17 & 0.14 \\ 
4311548412384207488* & 282.911 & +10.180 & 1.184 &  	& 1.030 & (0.7) & 3.22 & 0.08 \\ 
4311609920662656512* & 282.912 & +10.476 & 0.429 & 0.322  &  & 0.7 & 2.81 & 0.06 \\ 
4311553055261728256 & 282.913 & +10.271 & 1.366 &  	& 1.138 & 0.5 & 5.9 & 0.20 \\ 
\hline
4311549997275268352* & 282.921 & +10.245 & 1.219 &      & 1.029 & (0.8) & 6.92 & 0.18 \\
4311561095471630848* & 282.925 & +10.415 & 1.153 &  	& 0.950 & 0.8 & 2.53 & 0.05 \\ 
4311562435501392000 & 282.973 & +10.433 & 1.011 & 0.699  &  & 1.0 & 6.56 & 0.28 \\ 
\hline
\end{tabular}
\tablefoot{(1) ICRS coordinates. (2) Calculated using \cite{Gruner2020}. (3) Membership probability from \cite{CantatGaudin2018}. The two stars with probability 
values in parentheses
are not from the CG18 study but were instead calculated using 
the equation of \cite{Sagar1987}. Asterisks indicate possible binaries.}
\label{tab:periods}
\end{table*}

Each exposure time group was analyzed independently in 
\textsc{Daophot II}.
This allowed for confirmation of rotation periods among stars that
appeared in more than one exposure group.
Of the 210 cluster members mentioned in Section \ref{sec:membership},
we constructed lightcurves for 187 probable members.
Of these, 182 are FGK stars and 
5 are A stars.
From these, we find periods for 45 FGK cluster members and 
2 of the A stars. 
All rotation periods found and the \textit{Gaia} EDR3 designations
along with catalog and calculated colors are shown in Table 
\ref{tab:periods}.
Final periods were selected via the method outlined in 
\ref{sec:periodselection}.
Stars either showed agreement between different 
exposure times within a few percent, or the harmonic or subharmonic periods
produced a stronger signal at a different exposure. 
The phased light curves and results from the period analysis methods for all member stars are shown in the Appendix Figure \ref{fig:phased1}.
All possible binary stars with strong period signals are photometric, 
with RUWE values between 0.86 and 1.13.
In most cases, the selected period shows agreement between the PDM,
Clean, and GLS methods, but in a few cases the noise level is 
higher than the magnitude changes and the minima with the PDM methods 
are weak, and so 
we pick our periods from the CLEAN and GLS methods.
The SL method is primarily useful for confirming a period in the PDM 
of less than three days.
With each phased light curve, we subtract the mean. 

\subsection{The Color-Period Diagram}\label{sec:CPD}

\begin{figure}
    \centering
    \includegraphics[width=1\columnwidth]{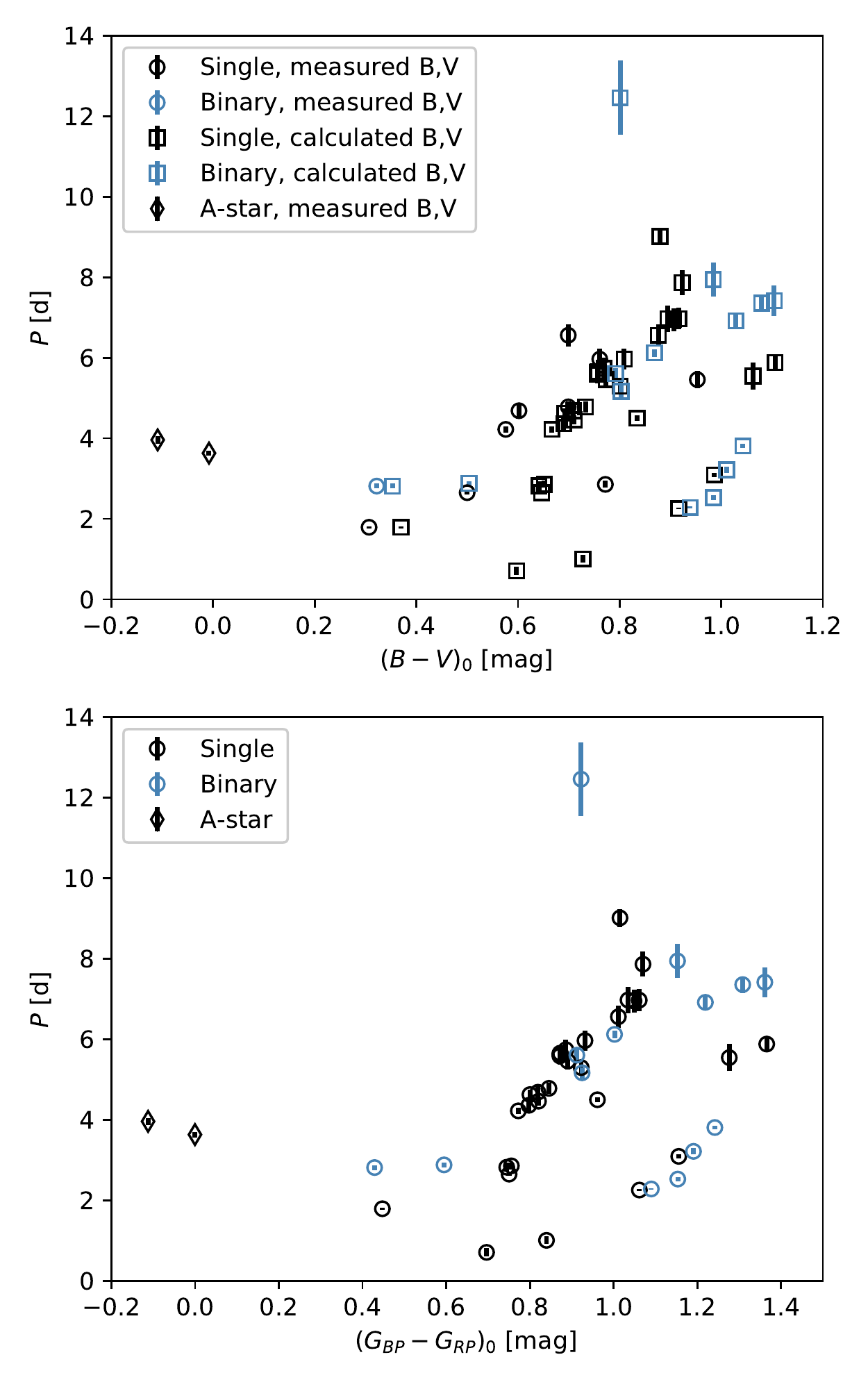}
    \caption{Rotation periods versus $(B-V)_0$ (top) and 
    $(G_{BP}-G_{RP})_0$ (bottom).
    Error bars are the $\sigma$-value of the Gaussian fit to the 
    minimum at the period in the PDM periodogram. Squares in $(B-V)_0$
    indicate magnitudes calculated from $(G_{BP}-G_{RP})_0$ while 
    circles are magnitudes found in the \textit{UCAC4} and 
    \textit{Gaia} EDR3 catalogs (top and bottom respectively).
    Possible binaries are indicated in blue.}
    \label{fig:CPD_basic}
\end{figure}

Our color-period diagram (hereafter CPD) is shown in 
Figure \ref{fig:CPD_basic} for both $(B-V)_0$ and 
$(G_{BP}-G_{RP})_0$.
We show the $(B-V)_0$ for comparison with other literature 
studies of rotation rates versus colors, but as can be seen here,
the sequence demonstrates more spread than for  $(G_{BP}-G_{RP})_0$
because the uncertainties in these colors are greater.
The CPD shows a sequence of increasing rotation periods towards 
redder stars. 
Below this sequence there is a small clump of rapidly rotating 
K-stars, which is to be expected for a cluster that is younger than  
$\approx 200$\,Myr. 
The photometric binaries lie below the slow rotator sequence, although 
these also have longer periods towards the cooler stars.
NGC 6709, due to observational limits, lacks observations for the M-dwarfs. 
Above the slow rotator sequence, there is at least one probable binary 
with a much slower rotation period and a membership probability of 
$0.6$.
There are also two bright A-type stars that had periodic brightening and dimming, but 
whether these are the result of being multi-star systems 
or some other phenomenon occurring in the line of sight is 
unknown. 
The rotation periods for these two stars only appeared in the 24\,s
exposure due to saturation at longer exposure times, but the 
periodic magnitude change is significant enough that a signal was 
found in multiple time series analysis methods.

\section{Discussion}\label{sec:discussion}

\begin{figure}
    \centering
    \includegraphics[width=1\columnwidth]{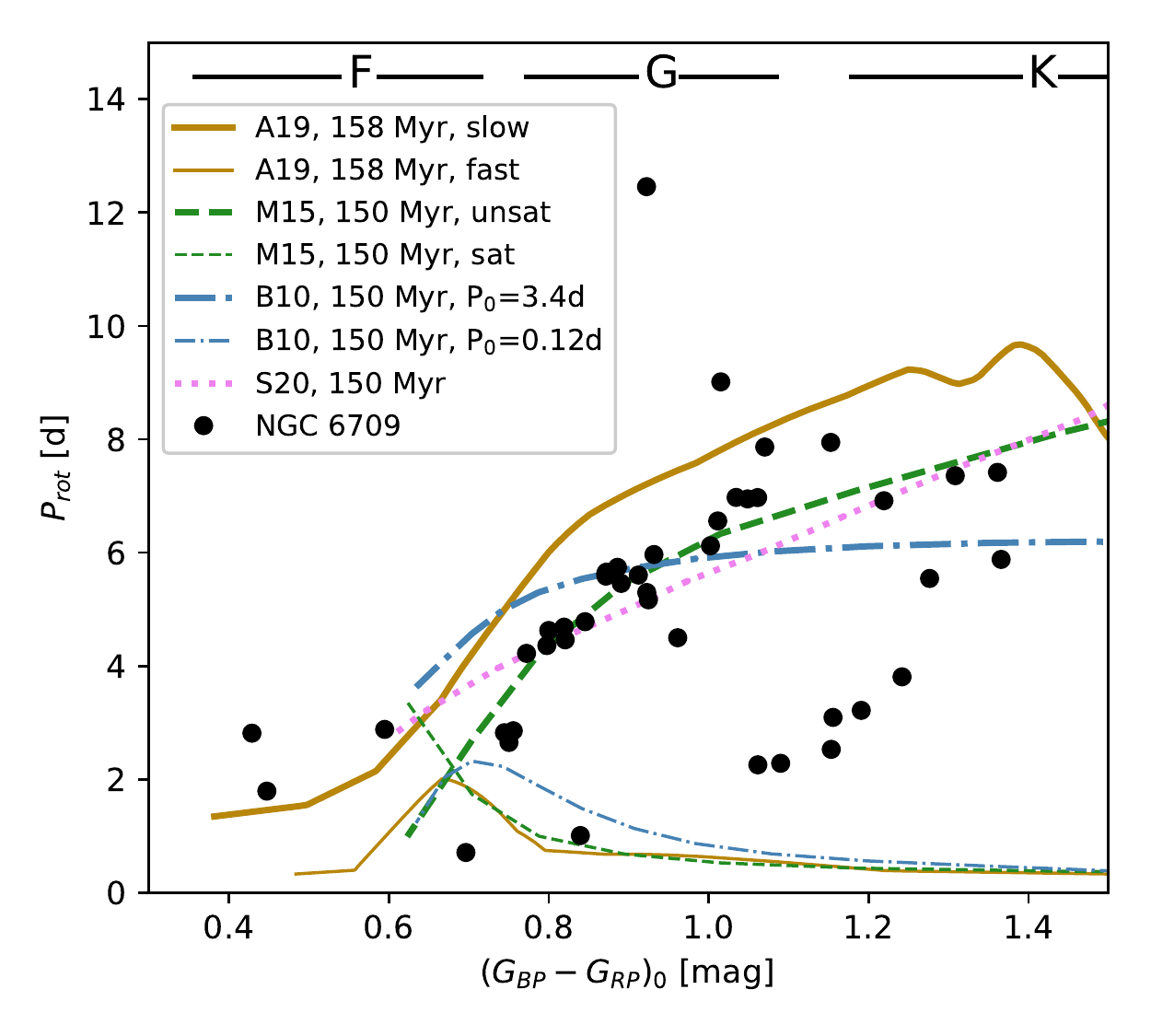}
    \caption{Rotation periods versus $(G_{BP}-G_{RP})_0$ 
    for cluster members and comparison to models.  
    A19 refers to the model by \cite{Amard2019}, where 
    both the sequences for initially slow and fast rotators are shown 
    by solid and dashed lines, respectively. M15 is the model of 
    \cite{Matt2015} with the unsaturated and saturated regimes 
    indicated by solid and dotted lines, respectively. 
    B10 is the model of \cite{Barnes2010a}, with the 
    solid and dashed lines again denoting initially slow and fast 
    rotators. S20 is the model of \cite{Spada2020}.}
    \label{fig:CPD_models}
\end{figure}

Due to the faintness of NGC\,6709, we use the
color system $G_{BP}$ and $G_{RP}$ from the \textit{Gaia} EDR3 
survey to avoid the necessity of converting the colors for
the fainter stars that lack observations in other passbands.
We use the 
conversion from \cite{Gruner2020} to convert between 
\textit{Gaia} colors and Johnson colors. 
The two bluest stars with rotation periods from NGC\,6709 
are omitted from the discussion as these 
are A-type stars and lack a convection zone, considered 
a critical component for the dynamo type that supports the 
formation of star spots \citep[e.g.,][]{Berdyugina2005}. 
Thus, their periodicity is likely the result of being members of 
a multi-star system and actual rotation rate is unknown.
For a more in-depth discussion of the behavior of these models 
including M-dwarfs, which are not present in our sample,
please see F20.

\subsection{Stellar spin-down models}\label{sec:gyromodels}

We compare the results with several models of stellar rotation
and age. 
These models are semi-empirical and use a theoretical framework 
built on the mechanisms thought to be responsible for 
stellar spin-down and are  calibrated to observations of, 
for example, the Sun, or well studied clusters such as the 
Hyades. 
We select the models of \cite{Barnes2010a} (hereafter B10), 
\cite{Matt2015} (hereafter M15),
\cite{Amard2019} (hereafter A19), and \cite{Spada2020} 
(hereafter S20) to derive 
theoretical isochrones for the estimated age of NGC\,6709.

The isochrones from the models are plotted against the rotation 
periods found for NGC\,6709 in Figure \ref{fig:CPD_models}.
All four stellar spin-down models have the same basic premise:
a ZAMS star with some initial rotation rate $P_0$
will begin to spin down due to coupling of the 
magnetic field with the stellar wind. 
The spindown rate depends on the Rossby number 
\begin{equation}
Ro = P_\mathrm{rot}/\tau_\mathrm{conv},
\end{equation}
where $\tau_\mathrm{conv}$ is the convective turnover time.
The selected models differ in some of the physical constraints and 
free parameters, but all reproduce to some degree the observed 
rotational behavior.

\subsubsection{The Barnes \& Kim 2010 model}\label{sec:barnes}

The B10 model is a dimensionless 
description of period evolution over time. 
The equation combines the evolution of both the fast (C)
and slow (I) sequences so that the period evolution equation is
\begin{equation}
    t = \frac{\tau_\mathrm{conv}}{k_C}\ln\left
    (\frac{P_\mathrm{rot}}{P_0}\right)+
    \frac{k_I}{2\tau_\mathrm{conv}}
    \left(P_\mathrm{rot}^2-P_0^2\right).
\end{equation}
where $t$ is the stellar age,  
$k_C$ and $k_I$ are dimensionless constants derived from observations, 
$P_\mathrm{rot}$ is the rotation period at stellar age, 
and $P_0$ is the initial rotation period.
In order to be consistent with previous work, $\tau_\mathrm{conv}$ is 
the global $\tau_c$ from 
Table 1 of \cite{Barnes2010a}, and the constants 
$k_C$ and $k_I$ are $0.646$\,d\,Myr$^{-1}$ and $452$\,Myr\,d$^{-1}$, respectively.
The initial rotation of the fast sequence is $P_0 = 0.12$\,d and for the 
slow sequence, $P_0 = 3.4$\,d.
These are based on observations of the extrema of initial rotation 
periods of young cluster members. 
The C and I sequences that emerge from these equations provide limiting 
cases.

The B10 model is shown in Figure \ref{fig:CPD_models} for an age of
150\,Myr with initial rotation periods of 3.4\,d and 0.12\,d 
as the solid and dotted blue lines respectively to recreate 
the limits of the I and C sequences.
This model results in rotation periods too slow for 
F-type stars and too fast for the K-type stars when compared to the 
observed rotation periods of NGC\,6709 for the I sequence. 
There are no cluster members with rotation periods shorter than the limit
created by the C sequence, which corresponds to stars with an initial
rotation period of 0.12\,d, which is near the break-up speed for 
FGK stars.
The rapid rotators are not well-populated, at least partially due 
to the lack of observations for M-stars.
\cite{Barnes2016} notes that this model 
does not account for evolution of the star pre-main sequence 
and works better for clusters and stars between 400\,Myr and 2.5\,Gyr.
The model appears to have too aggressive of a spin down for young stars 
that are solar or hotter, and is not aggressive enough for young stars 
that are cooler.
The G-type stars fit the model moderately well, which is not a surprise
as the constants for this model were calibrated using the solar age and 
rotation period.

\subsubsection{The Matt et al. 2015 model}\label{sec:matt}

The M15 model 
examines the changes in the torque over time on a star 
and the relationship with $Ro$.
Stellar activity is divided into two regimes, saturated and 
unsaturated, based on whether magnetic activity indicators 
have approached some saturated values independent of $Ro$.
The saturated and unsaturated regimes correspond to rapid and 
slow rotators, respectively.
We model the limit of the saturated and unsaturated regimes, 
Equations 10 and 11 respectively in M15, and input 
an age of 150\,Myr.
The moment of inertia is from the stellar tracks of 
\cite{Baraffe2015} and the convective turnover time is based on 
the empirical relation given in \cite{Cranmer2011}, as the 
model is calibrated to these values for $\tau$.
For the saturated limit, we use an initial rotation period 
of $P_0 = 0.2$\,d. 
Using the adopted parameters provided in Table 1 in the Erratum
which are adjusted to recreate solar values at a solar age,
we show the model in Figure \ref{fig:CPD_models}
with the solid line indicating the limit from the unsaturated regime
and the dotted line the limit from the saturated regime.  

The model provides a pretty good description of the G-stars and 
our few K-stars along the slow branch.
Our F-stars show rotation periods where the model predicts either very 
rapid rotation or none at all.  
The saturated limit for the torque is similar to the lower limit of the most 
rapid rotators from the B10 model, although it allows for slightly more 
rapid rotation in the solar regime. 
Few cluster members actually lie near this limit, but instead occupy the space 
between.

\subsubsection{The Amard et al. 2019 model}\label{sec:amard}

The rotation periods of A19 are based on 
models calculated using the STAREVOL v3.40 code.
The models follow structural and rotational evolution
and include the disc-coupling timescale and 
wind braking. 
To match our cluster, we use the models for 
solar metallicity ($Z = 0.0134$) and an age of 
$\log T = 8.2$, or 158\,Myr.
We show the models for both the slow and rapid rotators
in Figure \ref{fig:CPD_models} as solid and dotted gold lines respectively.
This corresponds to initial rotation rates of 
1.6 days (2.3 days for stars with $M > 1.2 M\odot$) 
for the rapid rotators and 9.0 days for the slow rotators. 

From the figure it can be seen that this model does not fit well, and 
a younger isochrone for the cluster would likely fit better as this one 
has an all-around too aggressive spin down for stars as compared to the 
other models.
This model, however, does reproduce better the rotation rates 
of our few F-type stars.
The lower limit for the rapid rotators matches fairly closely
with the corresponding saturated limit of the M15 model for the 
G- and K-type stars. 
This is not entirely surprising as the torque at the surface 
uses the M15 formulation for the change in angular momentum. 

\subsubsection{The Spada \& Lanzafame 2020 model}\label{sec:spada}

We also show the model of S20 
in Figure \ref{fig:CPD_models} in violet.
Unlike the previous models, this model has only the slow rotator sequence.
The model scales rotational coupling and stellar wind braking with 
mass to explain the lack of low-mass stars that have reached the 
slow rotator sequence for clusters younger than 1\,Gyr. 
We use the isochrones from Table A1 in the work of S10 
for 150\,Myr and the 
relation from \cite{Gruner2020} to convert $(B-V)_0$ to 
$(G_{BP}-G_{RP})_0$.

This model does reproduce fairly well the rotation periods 
for the K-type stars and the hotter G-type stars, 
but the spindown is not aggressive enough for  
the solar-type and cooler stars and the resulting 
rotation periods are too rapid 
compared to what has been observed for our cluster members. 

\subsection{Comparison to other open clusters}\label{sec:comparison}

\begin{figure}
    \centering
    \includegraphics[width=1\columnwidth]{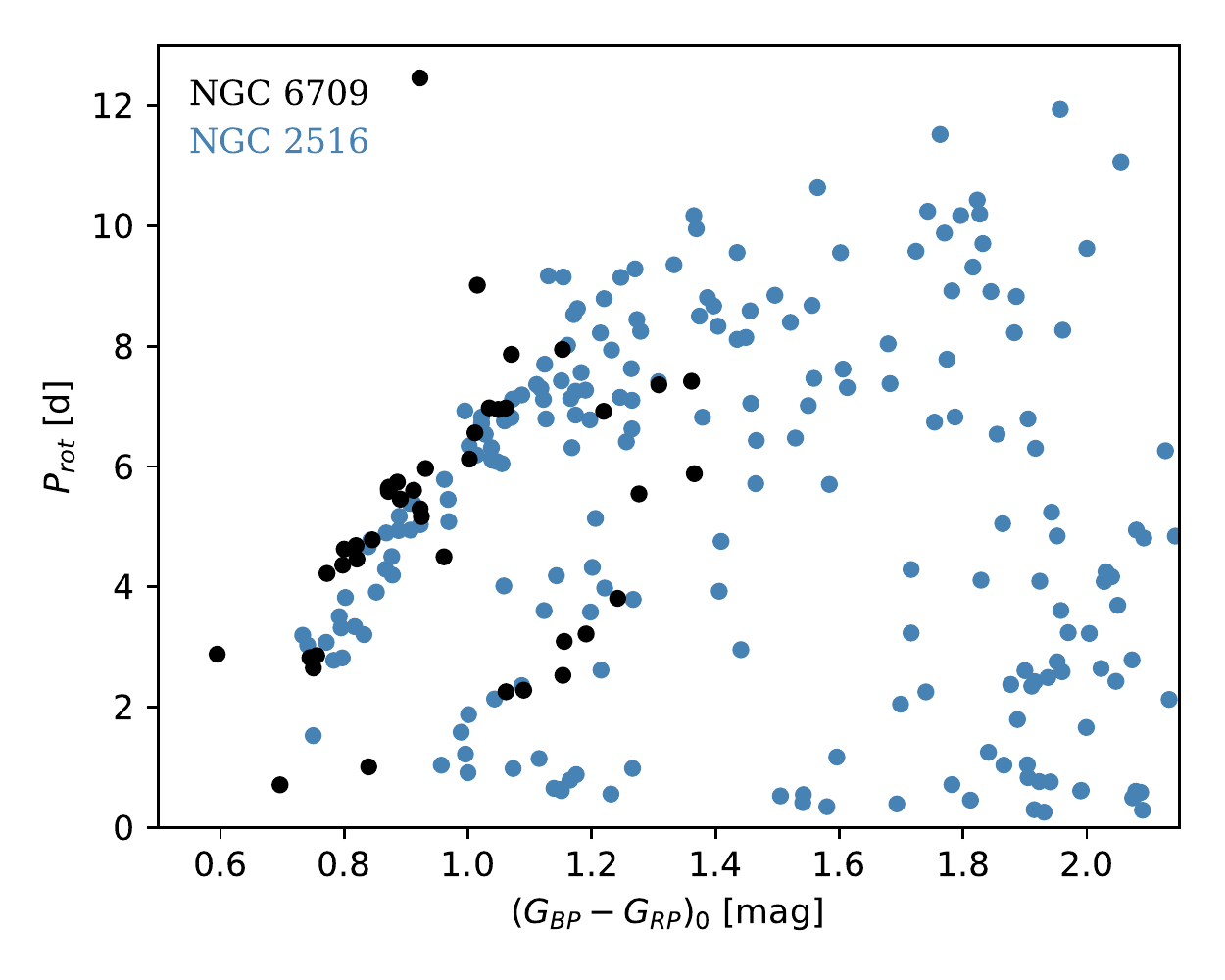}
    \caption{Overlay of the CPD of NGC 6709 on that of NGC 2516 from 
    F20. The two distributions appear to be identical in the region of overlap.}
    \label{fig:CPDcomp}
\end{figure}

We compare the CPD of NGC\,6709 to the similarly aged clusters 
detailed in F20.
It can be seen in Figure \ref{fig:CPDcomp} from overlaying 
NGC\,6709 with NGC\,2516, the benchmark cluster from that work,
that although NGC\,6709 is fainter and has a 
few giant members, it exhibits almost identical rotational behavior. 
F20 also compare their open cluster with other similarly aged open clusters 
M35, the Pleiades, M50, and Blanco I in their Figure 17,
and note the near identical structure of the rotation periods versus
color. 
Our cluster is no exception, and fits in well with the age group of 
Pleiades-like open clusters. 

The slow rotator sequence of NGC\,6709 is very similar to that of 
NGC\,2516. 
The slow-sequence for NGC\,6709 is very slightly above that of NGC\,2516,
and so this cluster should accordingly be slightly older. 
There is an outlier with a much longer rotation period than expected
($\approx 12.5$\,d),
but similar outliers are observed for M35, the Pleiades, and 
M50. 
NGC\,6709 also exhibits the gap between stars on the rapid 
rotator sequence and stars that have evolved onto the slow 
rotator sequence. 
The rapid rotators for NGC\,6709 all have color 
values $(G_{BP}-G_{RP})_0 < 1.5$, and so a comparison of 
the structure of the rapid rotators that have not yet joined the 
slow rotator sequence 
is not possible.

\section{Conclusions}\label{sec:conclusions} 

We observe the open cluster NGC\,6709 over multiple observing seasons 
to obtain for the first time 2--4 month long photometric light curves
using \textsc{Daophot II}.
Member FGK stars are identified and 
analyzed using four different methods
(PDM, SL, GLS, and CLEAN)
to find periodicities.
By comparing the results from these four methods, we obtain 
rotation periods for 47 member stars within our field. 
We calculate the uncertainty of these rotation periods from 
the FWHM of the peak from the PDM method.

Using \textit{Gaia} colors, 
we construct a color-period diagram of cluster members.
This CPD has a sequence of slow rotators with increasing rotation
periods towards 
the redder stars and a small clump of rapidly rotating stars 
that have not yet converged onto the aforementioned slow sequence.
The six photometric binary members show no significant difference and 
follow the same pattern as other member stars.
We compare isochrones from theoretical, empirical, and semi-empirical  
models for stellar rotation and evolution to our CPD
and estimate the cluster age to be about 150\,Myr. 
This is within the range of ages estimated from literature based
on isochrones from color-magnitude comparisons.
The models that best fit the CPD of NGC\,6709 are those
of M15 and S20.

The rotation periods follow a similar behavior as that observed
in other open clusters of similar ages and characteristics. 
We compare this cluster's CPD to NGC\,2516 in the work of 
F20 and find that the rotation period distribution 
for NGC 6709 matches that of NGC 2516.
F20 also compares NGC\,2516 to four more open clusters of similar ages,
the Pleiades, M35, M50, and Blanco I.
This is further support that the mechanisms driving the star formation 
and subsequent spin-down 
for stars with similar parameters are the same for open clusters 
with sufficiently similar parameters of age and metallicity.

\begin{acknowledgements}
ECK gratefully acknowledges funding from the the Deutsche Forschungs-Gemeinschaft (DFG project 4535/1-1B). 
STELLA was made possible by funding through the State of Brandenburg (MWFK) and the German Federal Ministry of Education and Research (BMBF) and is run collaboratively with the IAC in Tenerife, Spain. 
This work has made use of data from the European Space Agency (ESA) mission
{\it Gaia} (\url{https://www.cosmos.esa.int/gaia}), processed by the {\it Gaia}
Data Processing and Analysis Consortium (DPAC,
\url{https://www.cosmos.esa.int/web/gaia/dpac/consortium}). Funding for the DPAC
has been provided by national institutions, in particular the institutions
participating in the {\it Gaia} Multilateral Agreement.
This research made use of Astropy,\footnote{\url{https://www.astropy.org}} a community-developed core Python package for Astronomy.

 \end{acknowledgements}

\bibliographystyle{aa}
\bibliography{NGC6709.bib}
\begin{appendix}
\onecolumn
\section{Phased light curves for FGK member stars with rotation periods}
\begin{figure}[H]
     \centering
     \includegraphics[width=0.3\linewidth]{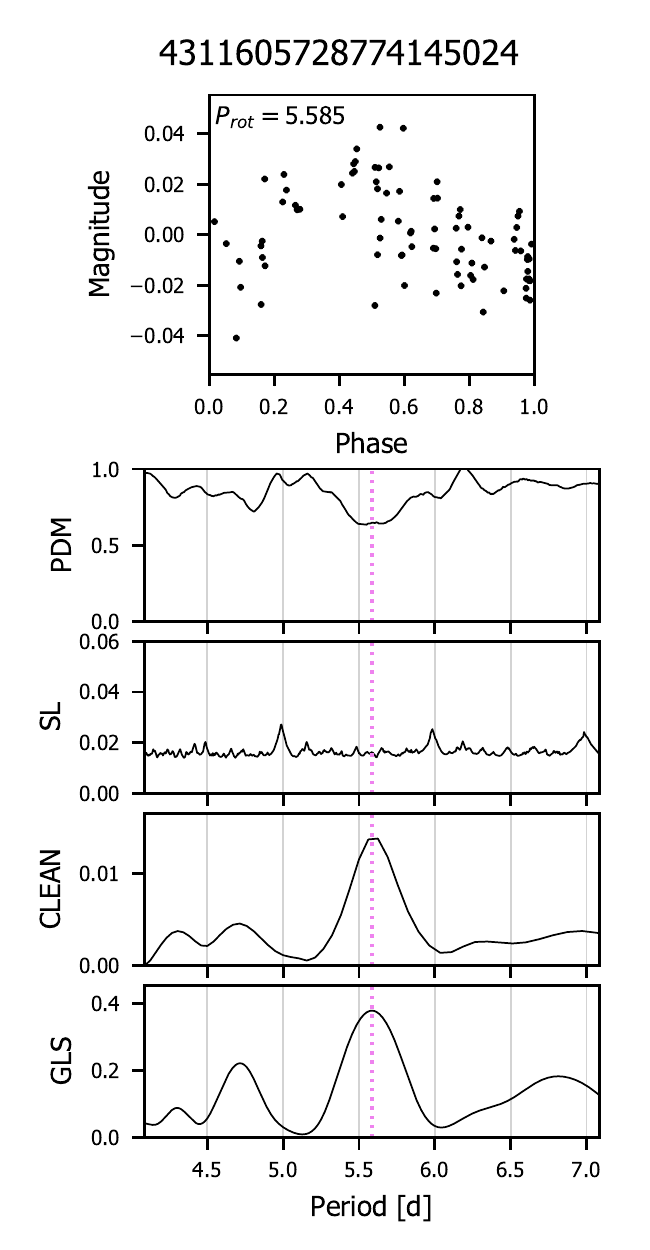}
     \includegraphics[width=0.3\linewidth]{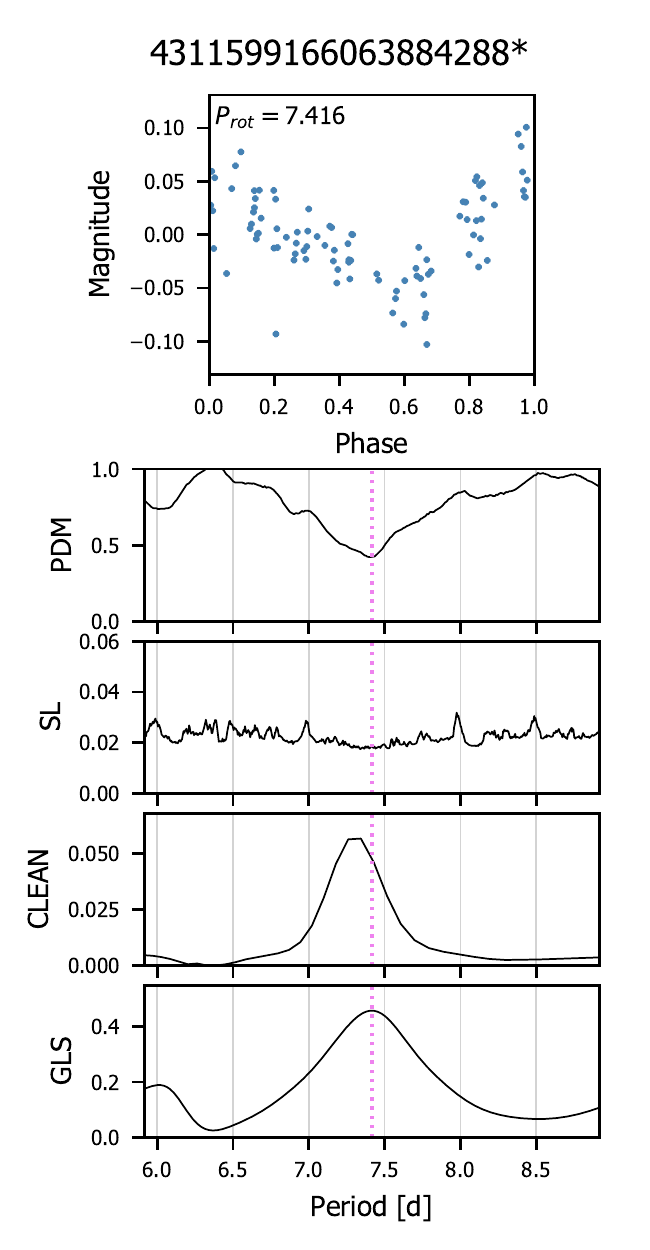}
     \includegraphics[width=0.3\linewidth]{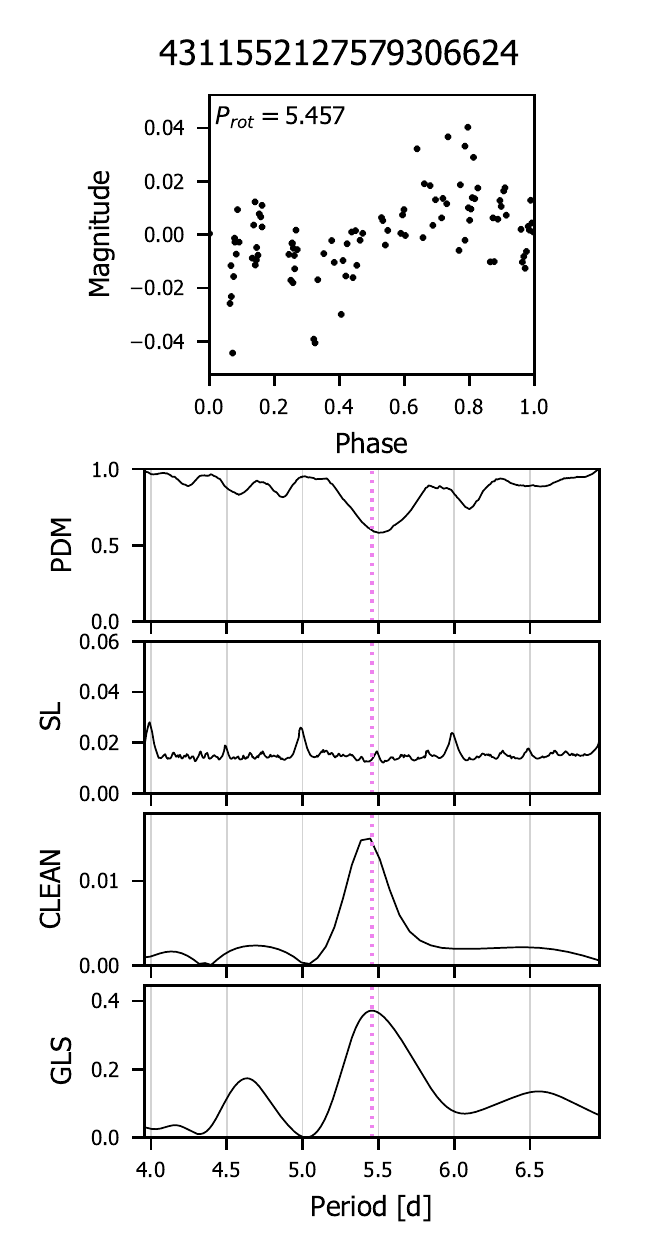}
     \includegraphics[width=0.3\linewidth]{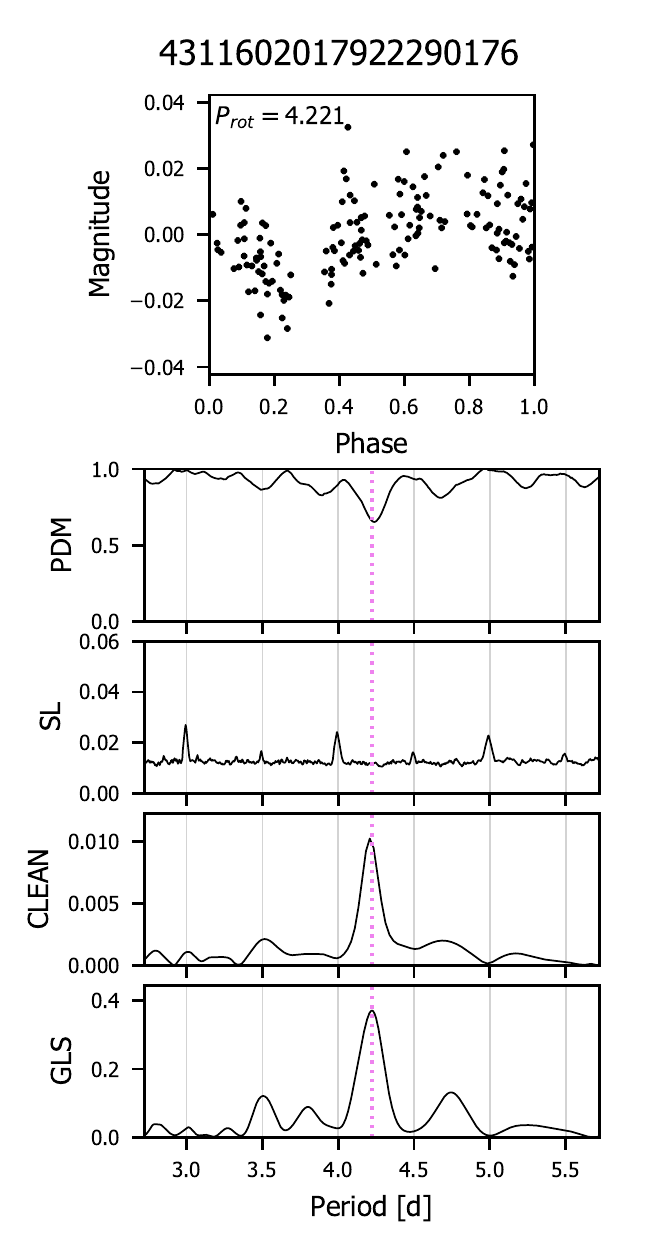}
     \includegraphics[width=0.3\linewidth]{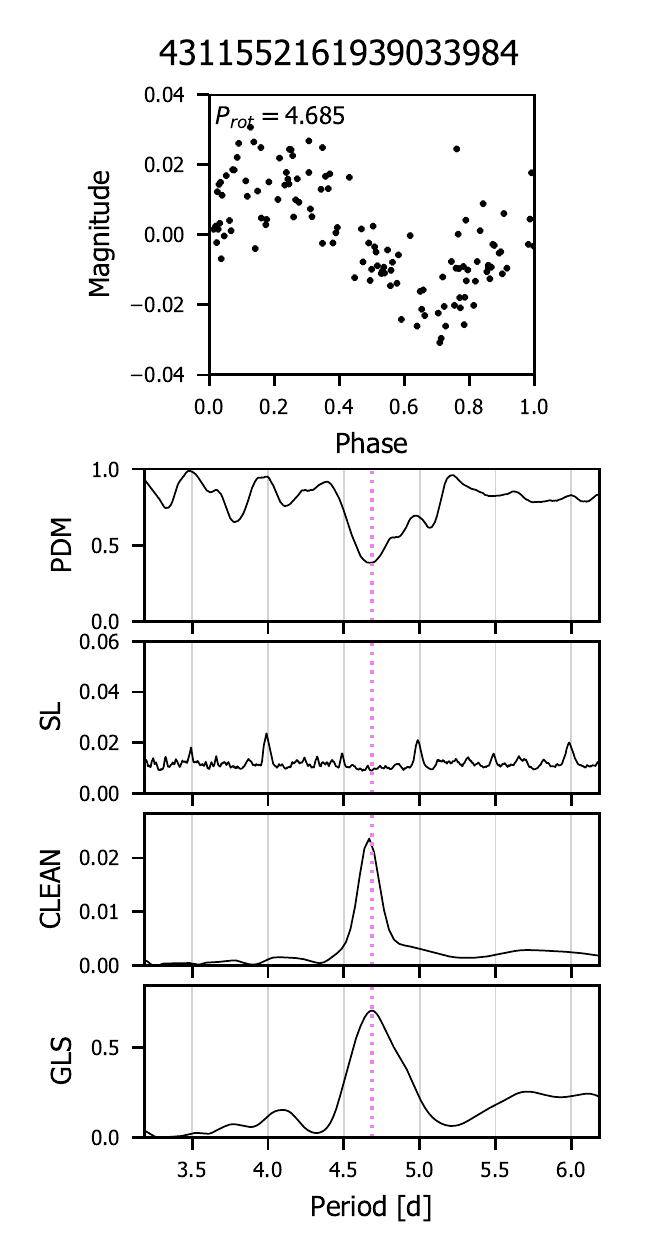}
     \includegraphics[width=0.3\linewidth]{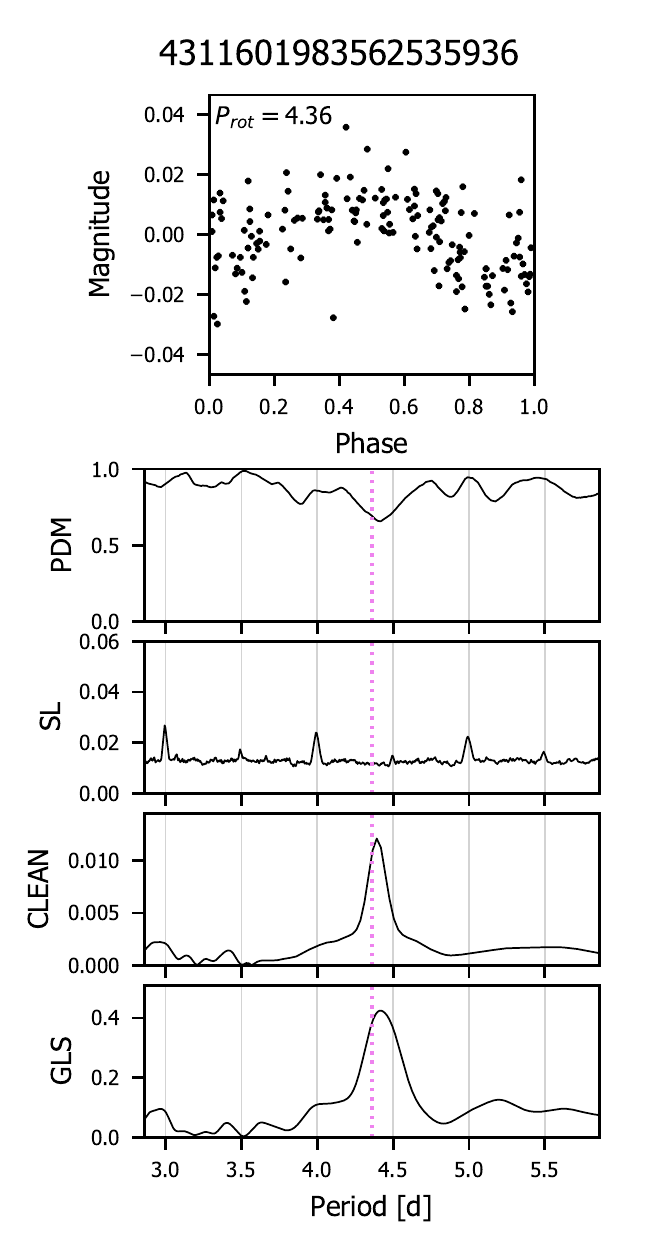}
     \caption{Phased light curves and results of period analysis methods. 
     An asterisk after the \textit{Gaia} designation and blue phased light curve indicates a possible binary. The vertical dotted line marks
     the found period.}
     \label{fig:phased1}
\end{figure}

\begin{figure}\ContinuedFloat
     \centering
     \includegraphics[width=0.3\linewidth]{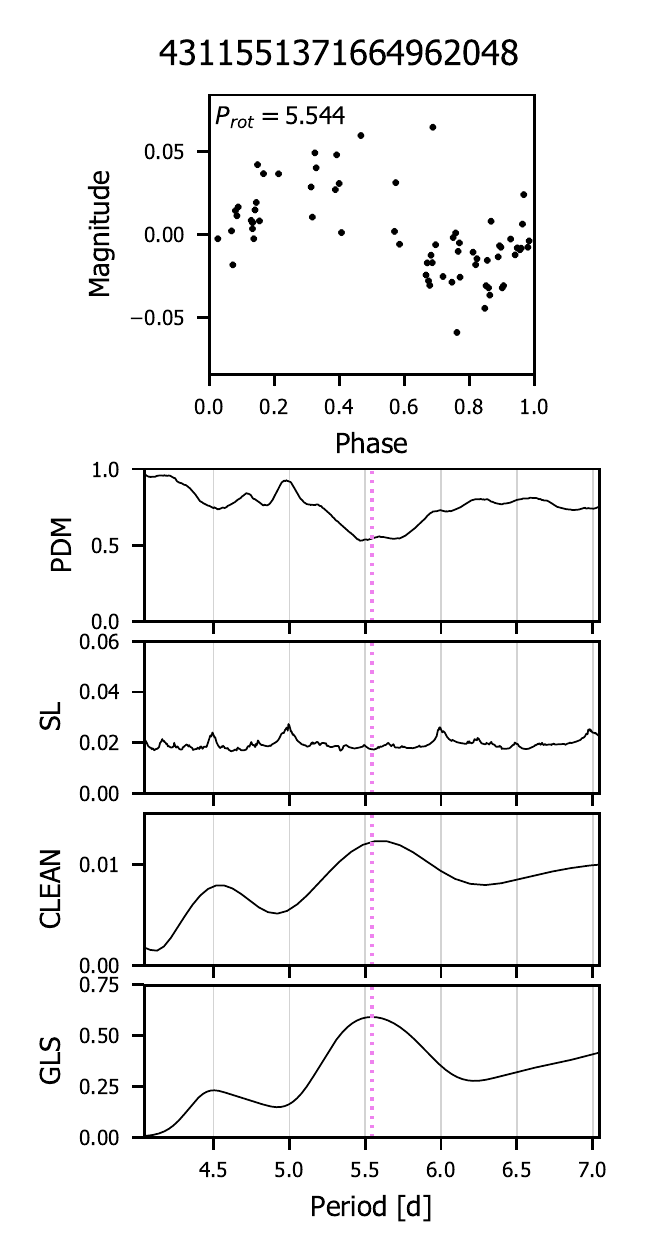}
     \includegraphics[width=0.3\linewidth]{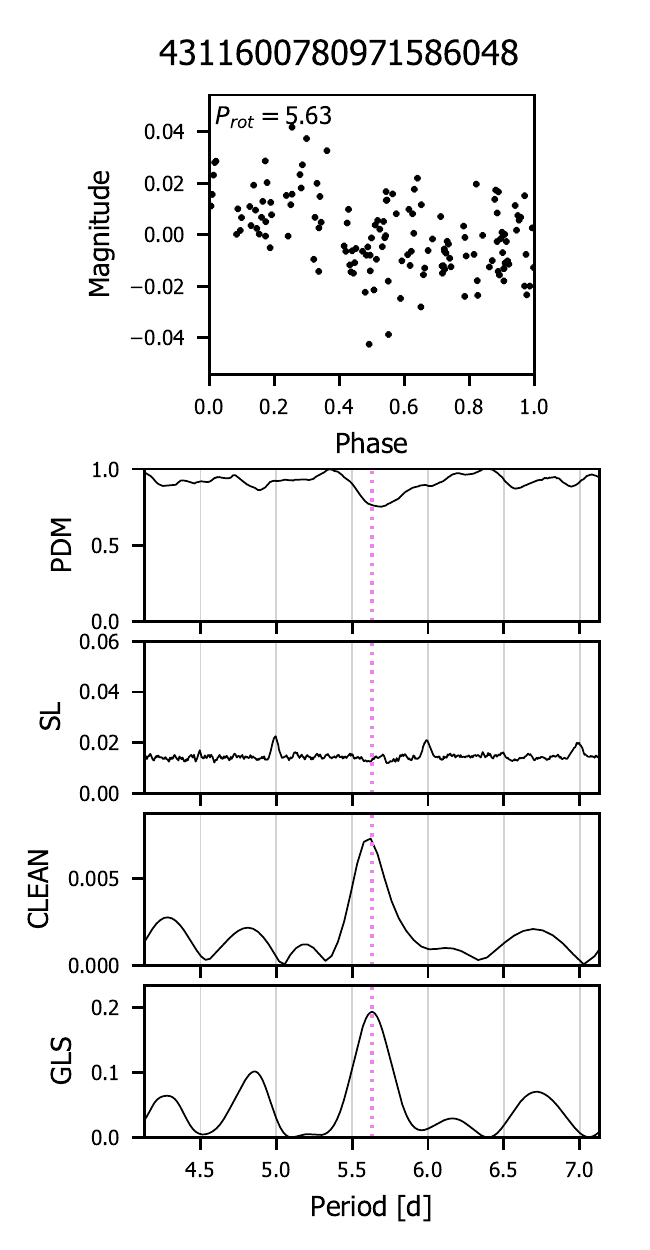}
     \includegraphics[width=0.3\linewidth]{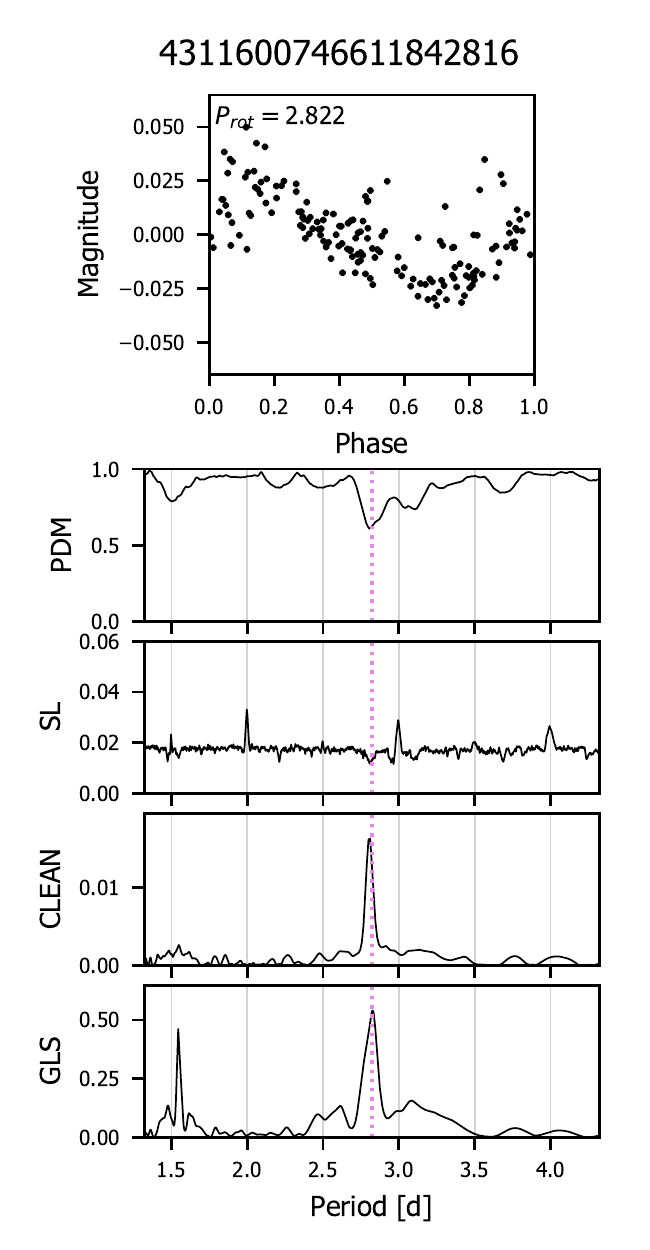}
     \includegraphics[width=0.3\linewidth]{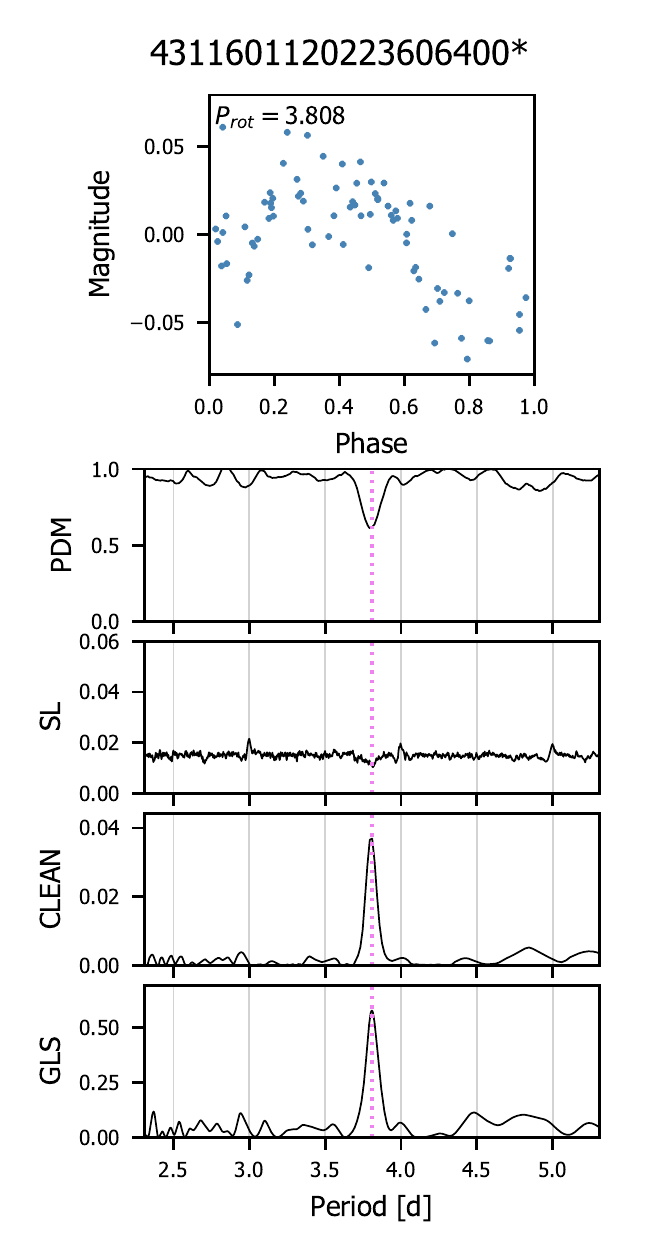}
     \includegraphics[width=0.3\linewidth]{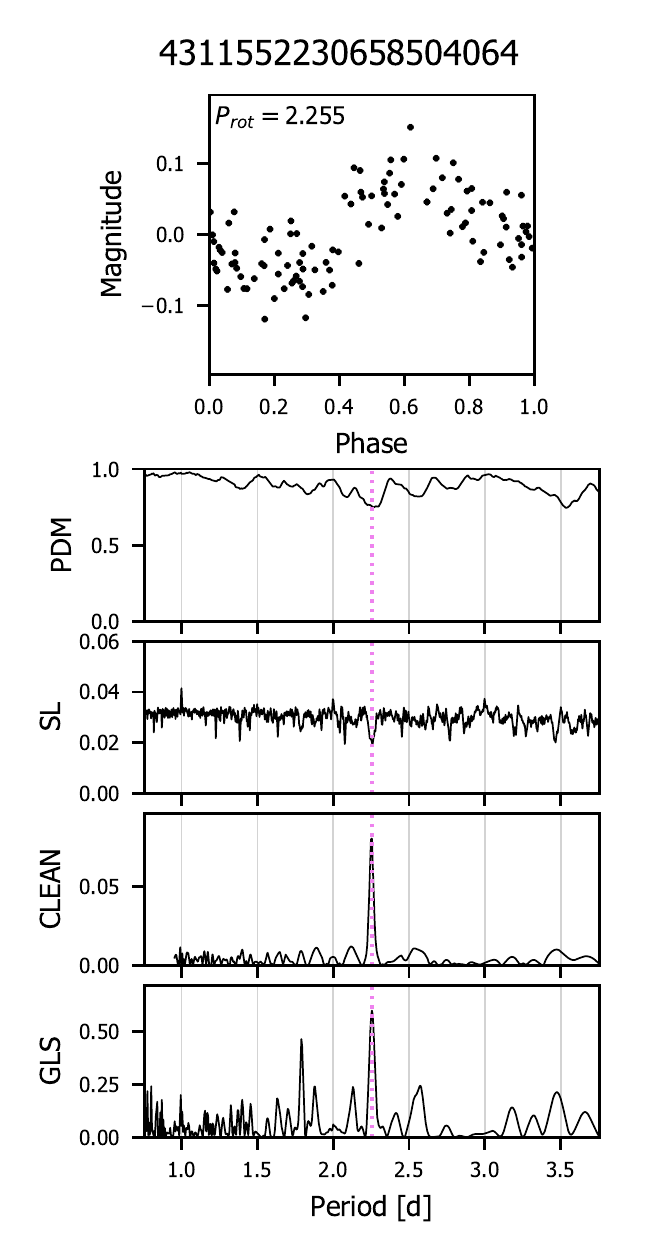}
     \includegraphics[width=0.3\linewidth]{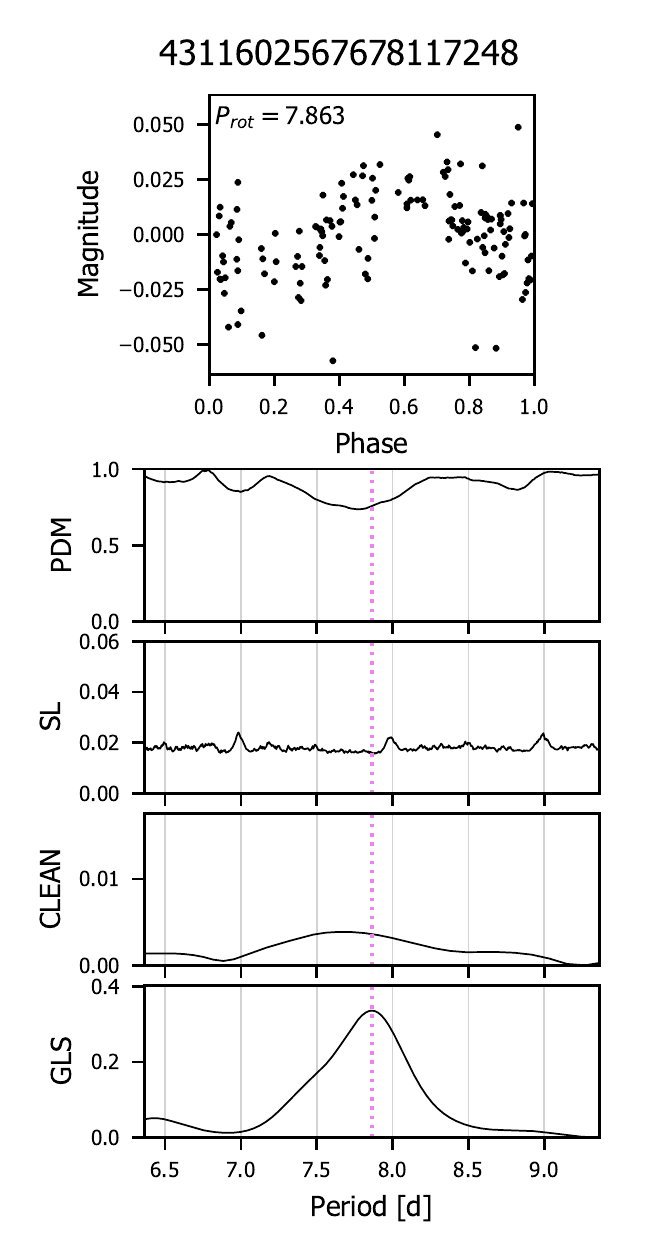}
     \caption{continued.}
     \label{fig:phased2}
\end{figure}

\begin{figure}\ContinuedFloat
     \centering
     \includegraphics[width=0.3\linewidth]{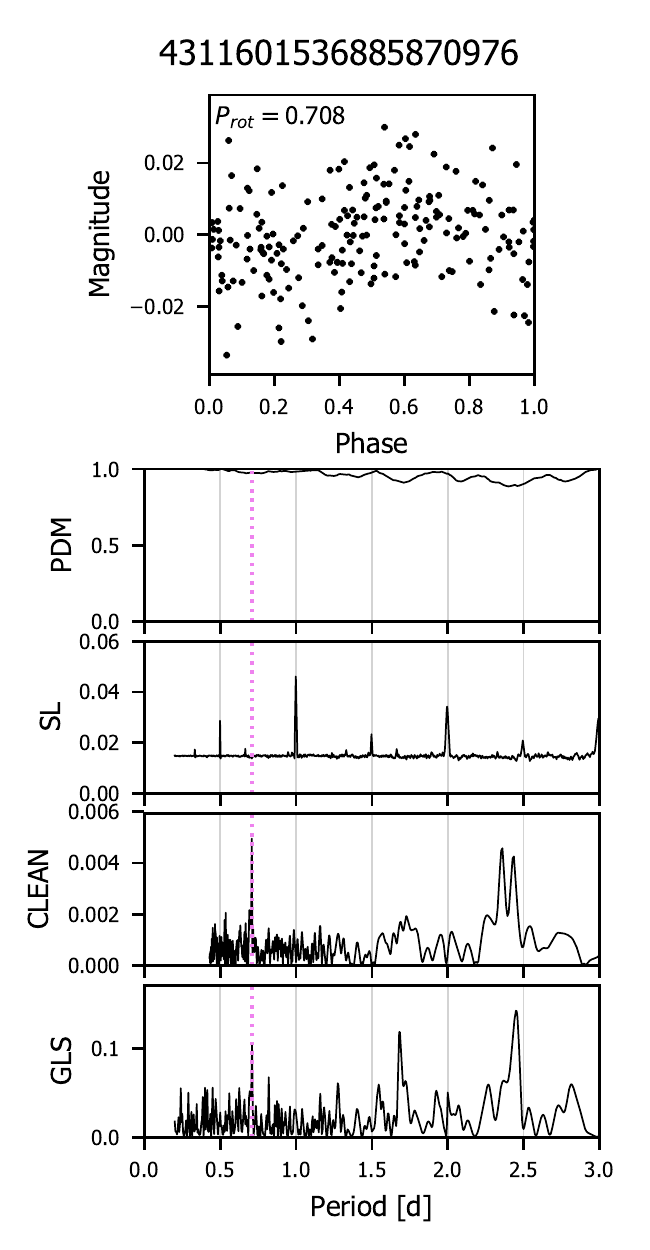}
     \includegraphics[width=0.3\linewidth]{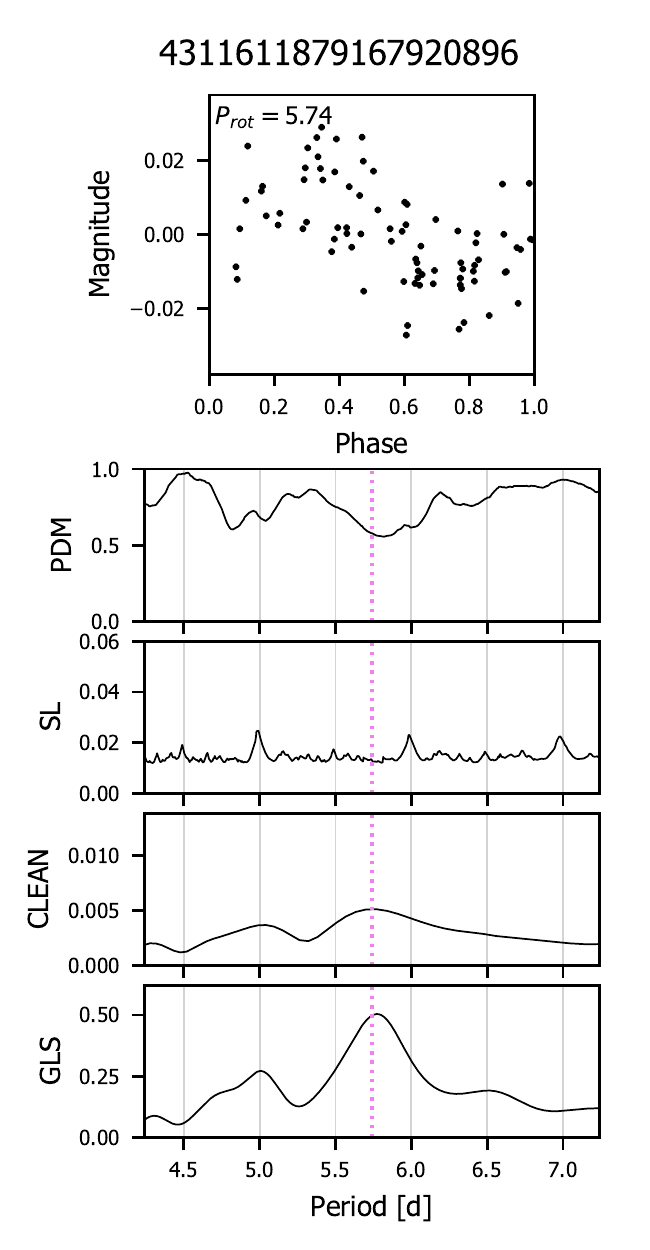}
     \includegraphics[width=0.3\linewidth]{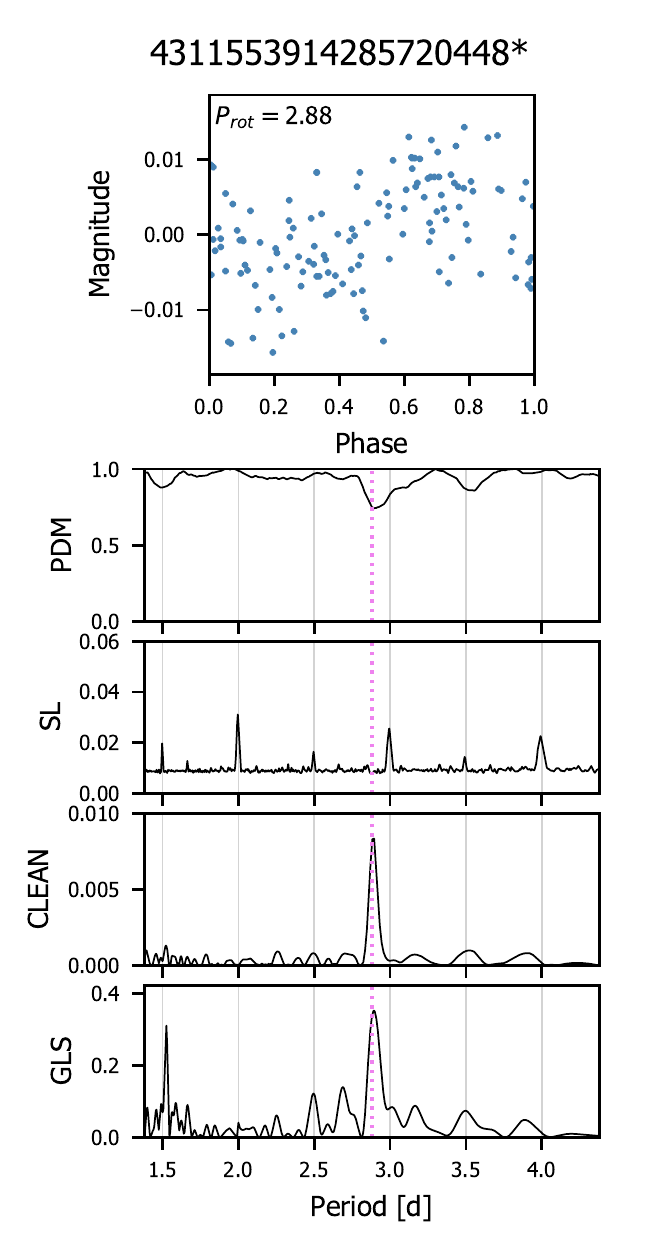}
     \includegraphics[width=0.3\linewidth]{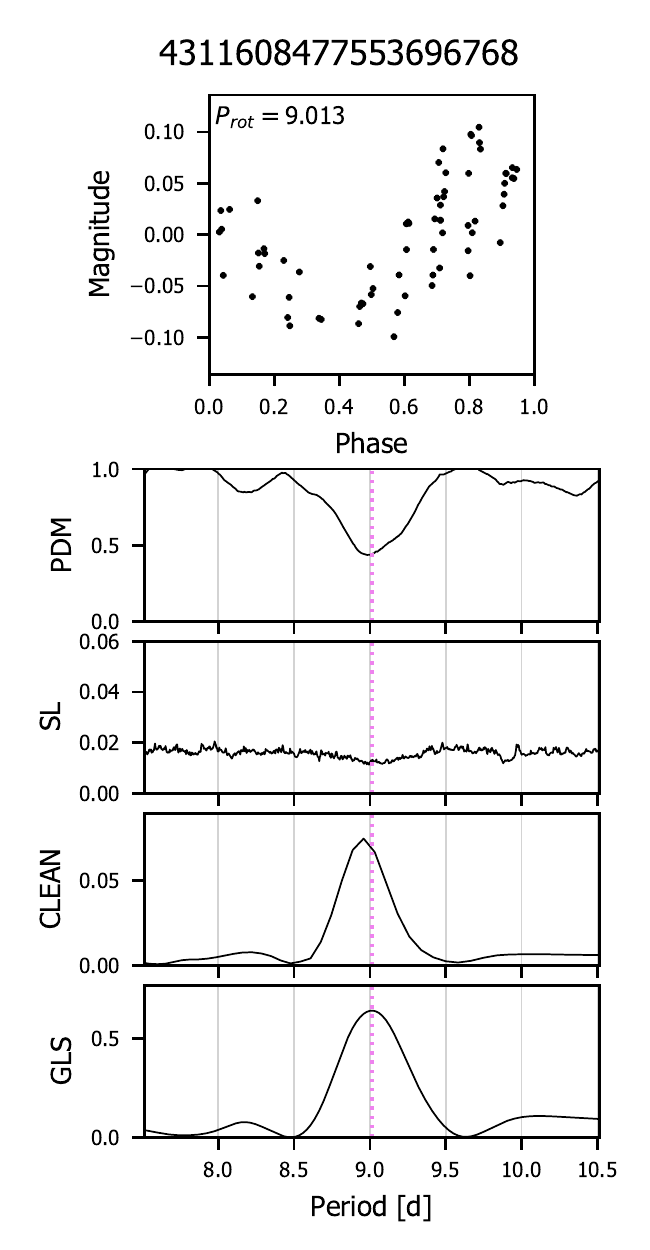}
     \includegraphics[width=0.3\linewidth]{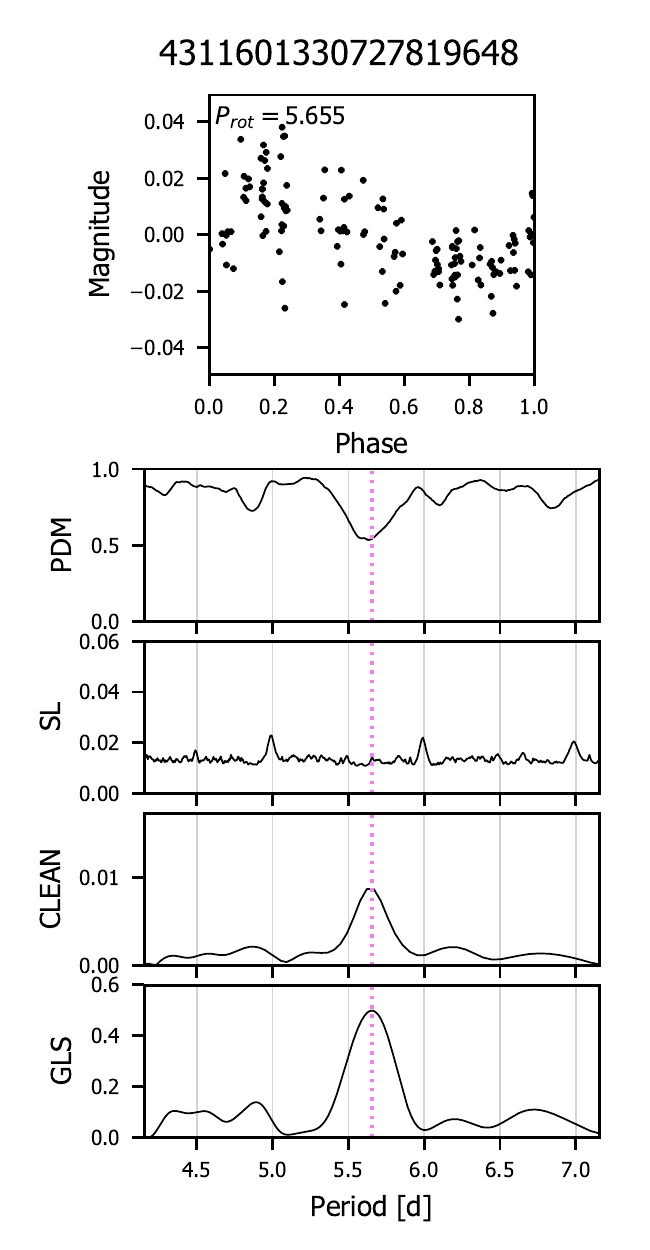}
     \includegraphics[width=0.3\linewidth]{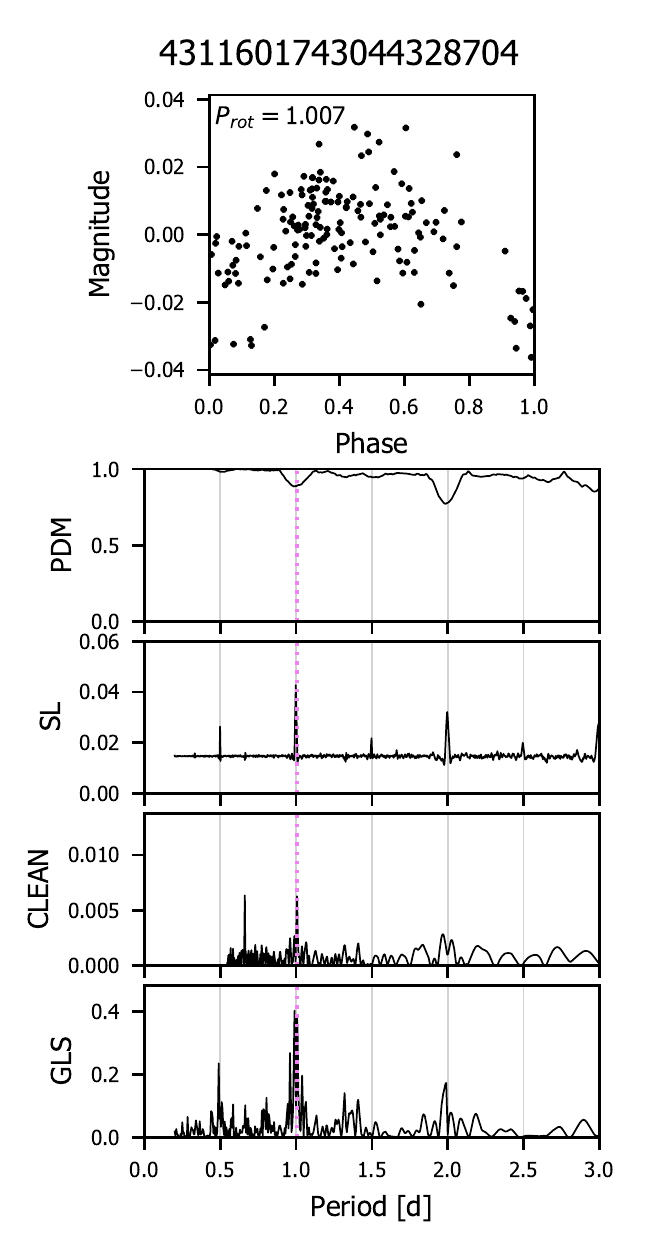}
     \caption{continued.}
     \label{fig:phased3}
\end{figure}

\begin{figure}\ContinuedFloat
     \centering
     \includegraphics[width=0.3\linewidth]{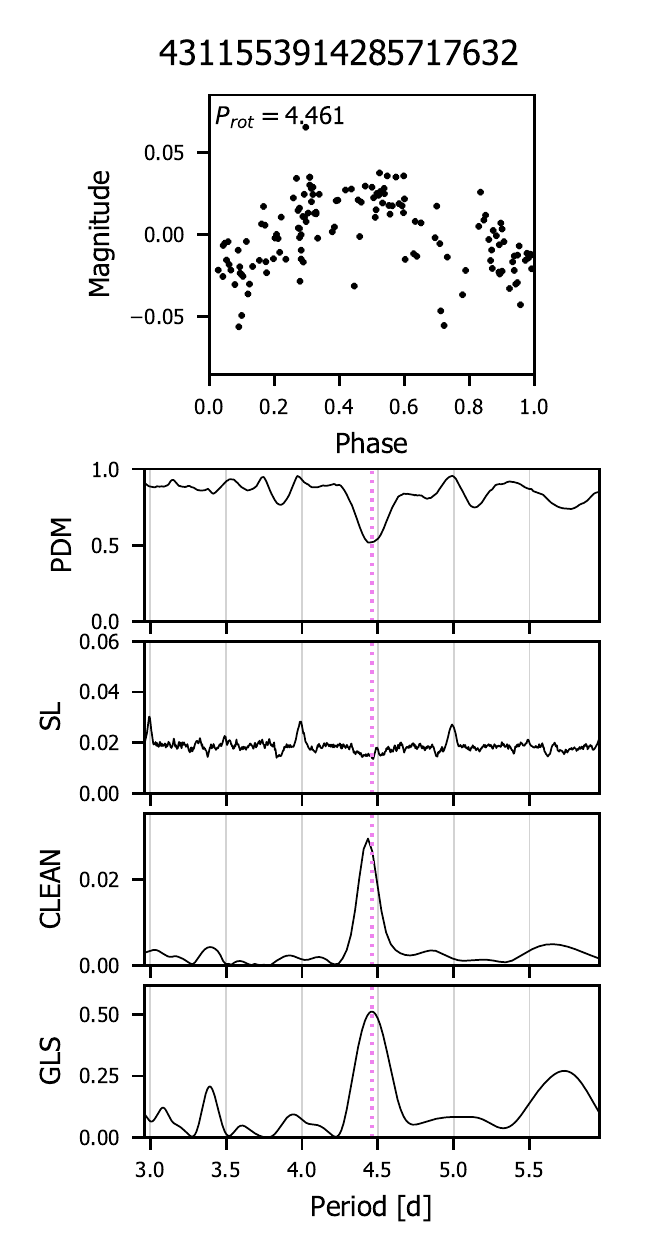}
     \includegraphics[width=0.3\linewidth]{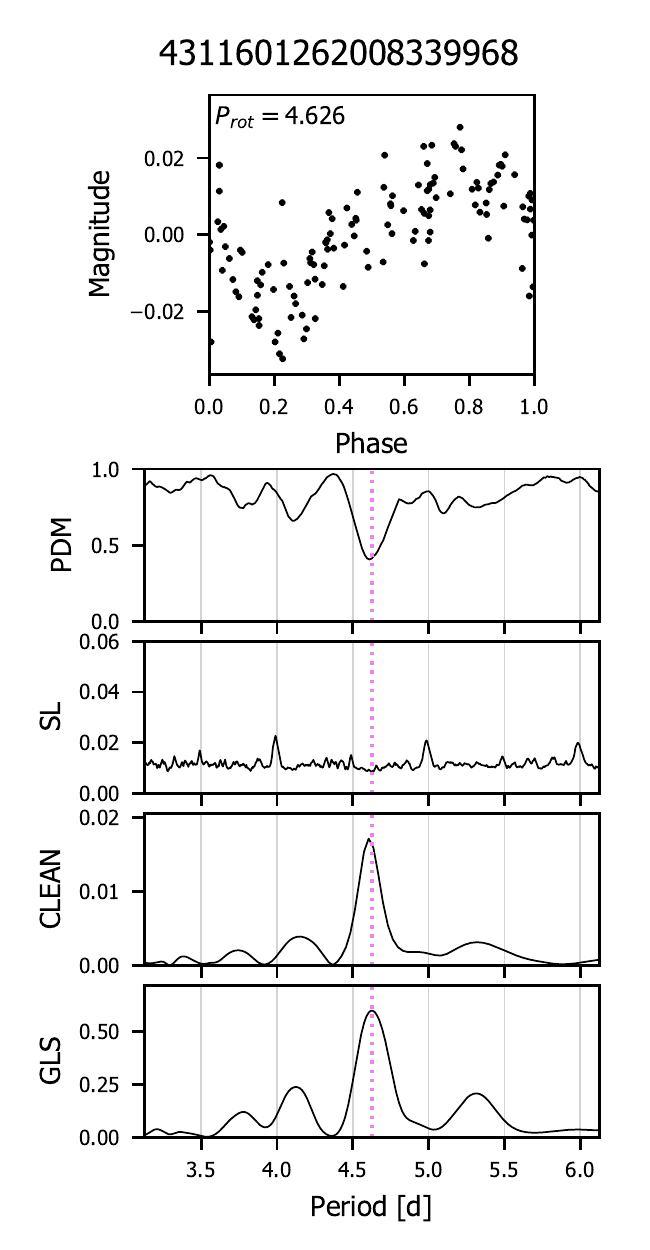}
     \includegraphics[width=0.3\linewidth]{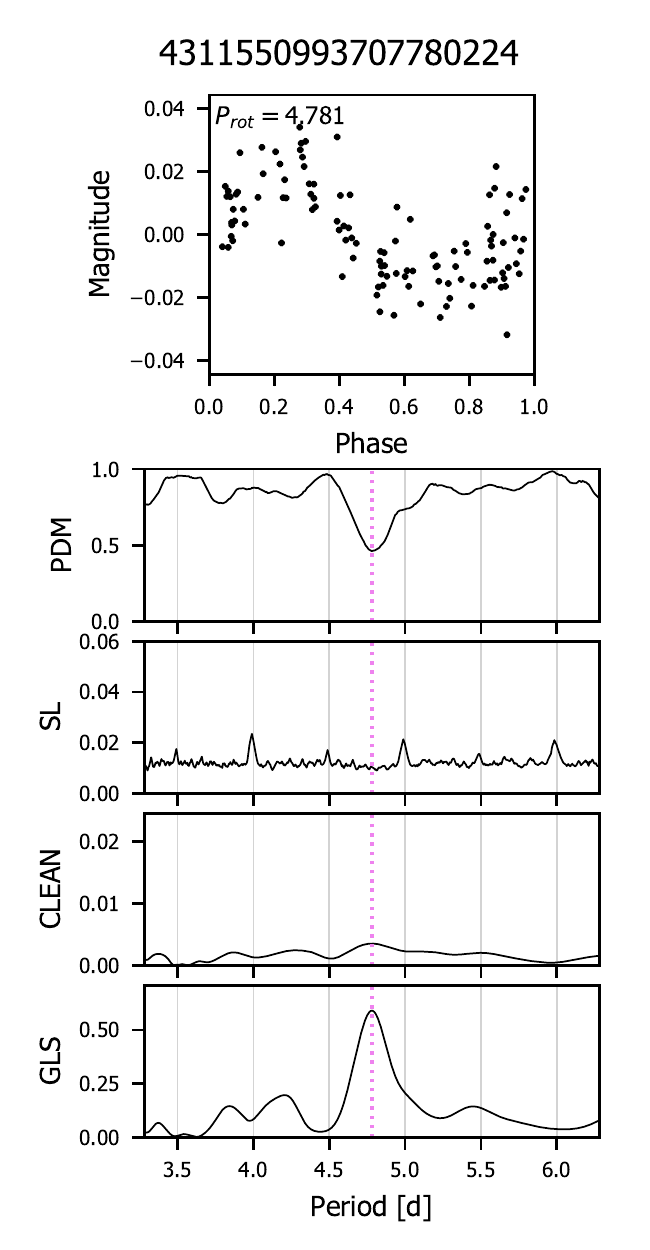}
     \includegraphics[width=0.3\linewidth]{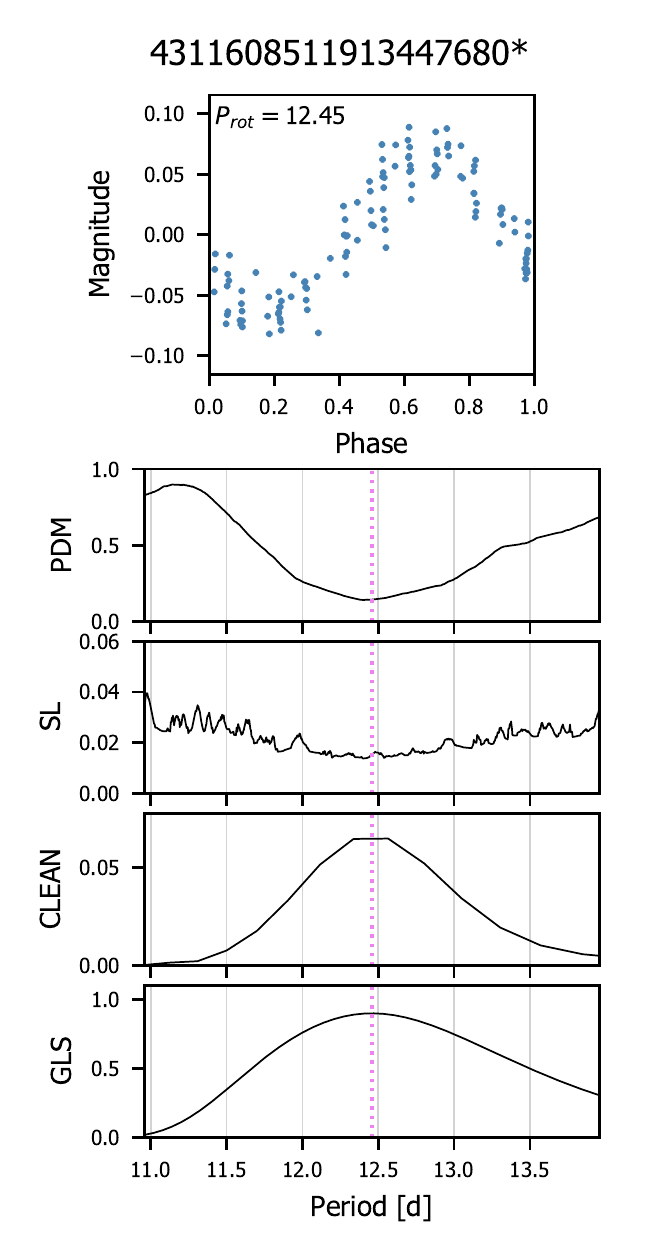}
     \includegraphics[width=0.3\linewidth]{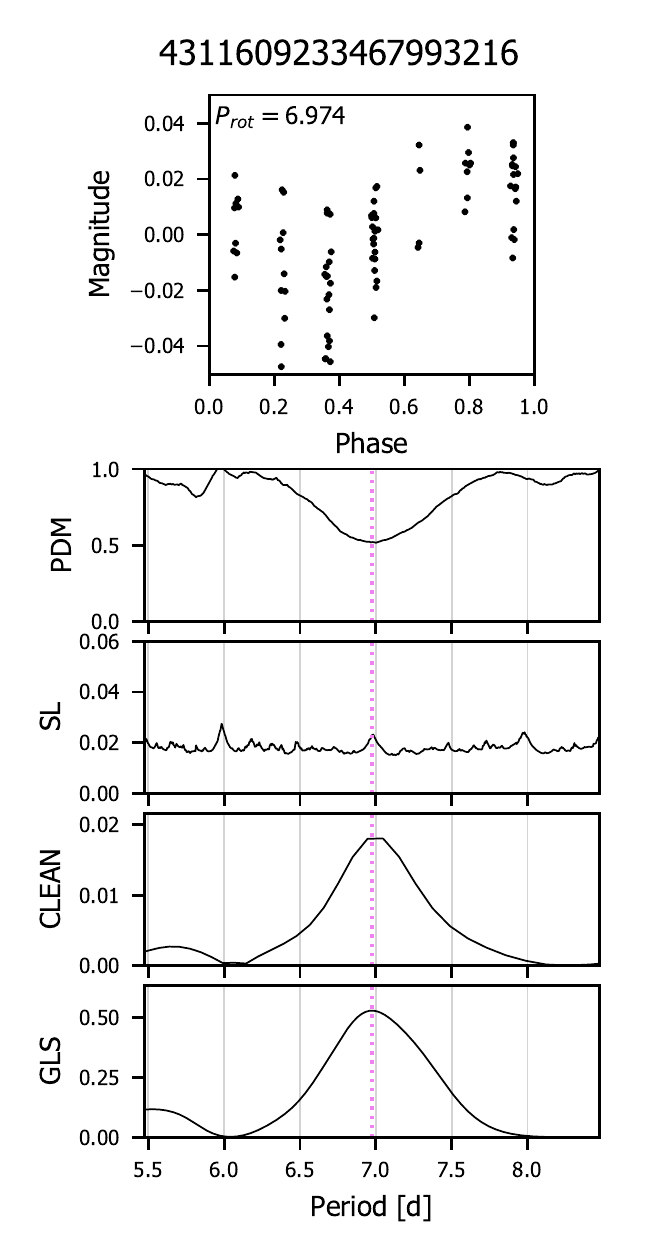}
     \includegraphics[width=0.3\linewidth]{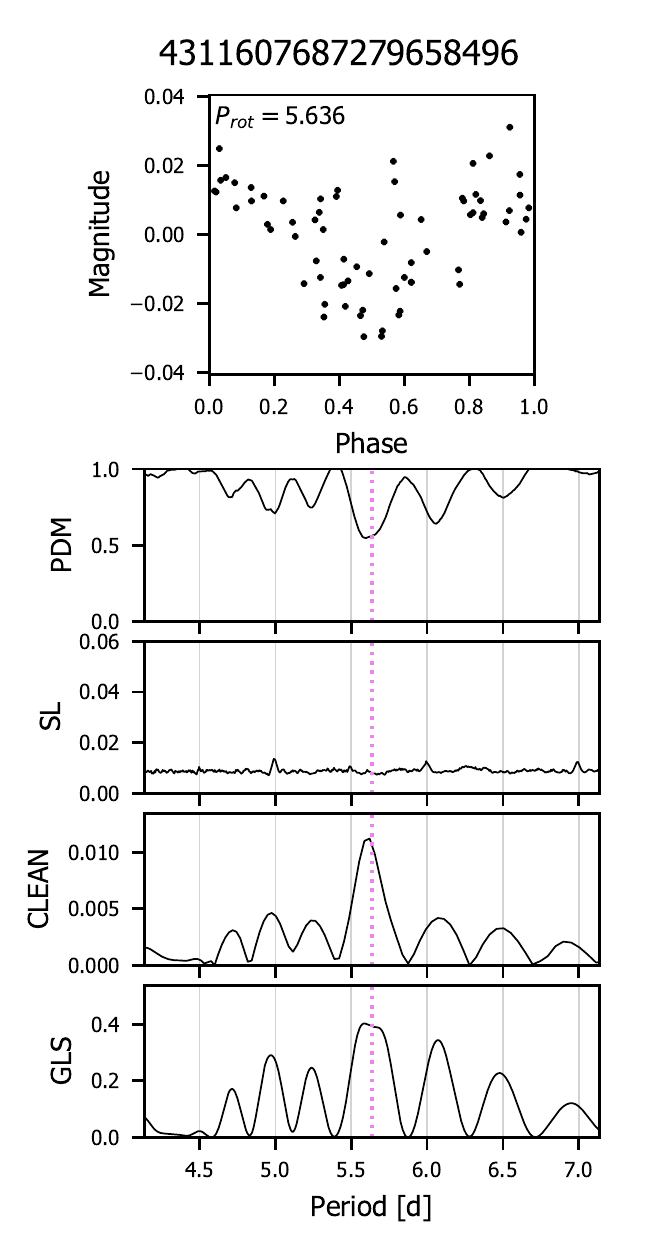}
     \caption{continued}
     \label{fig:phased5}
\end{figure}

\begin{figure}\ContinuedFloat
     \centering
     \includegraphics[width=0.3\linewidth]{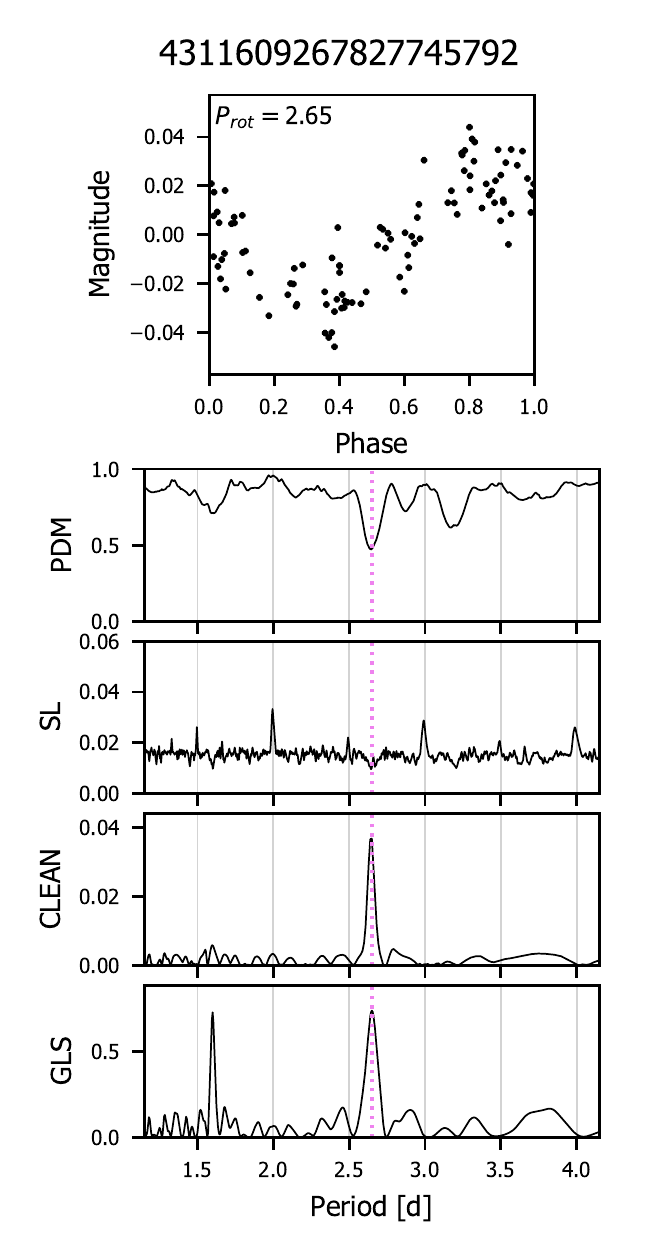}
     \includegraphics[width=0.3\linewidth]{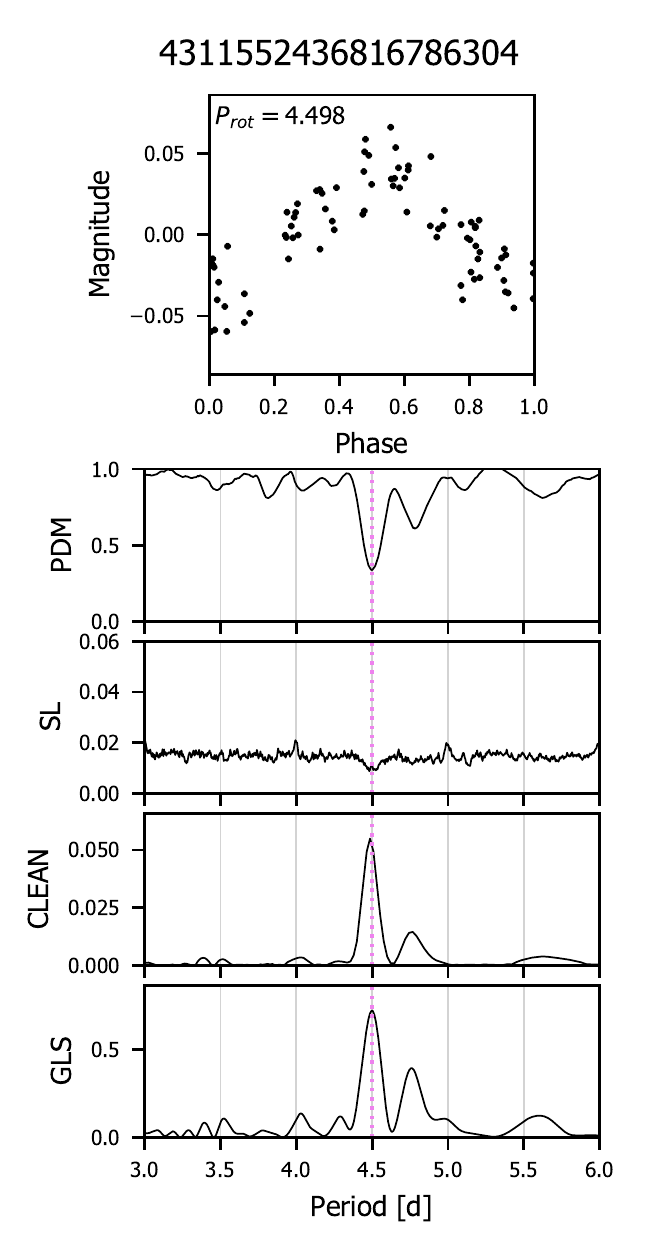}
     \includegraphics[width=0.3\linewidth]{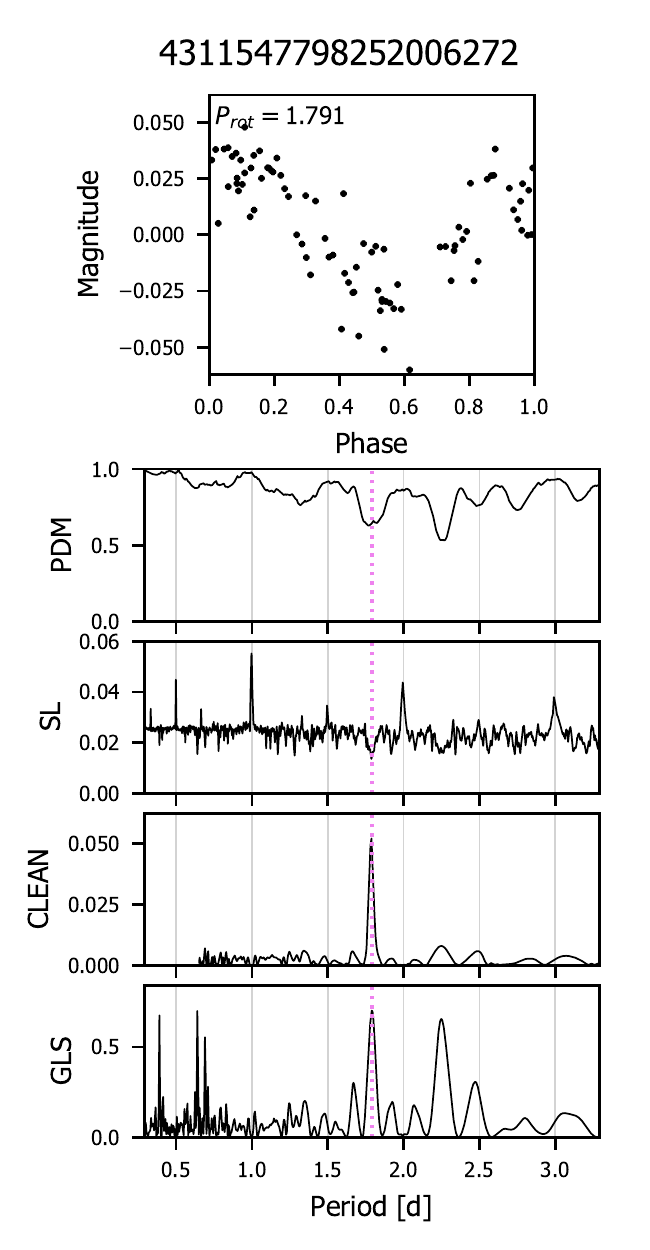}
     \includegraphics[width=0.3\linewidth]{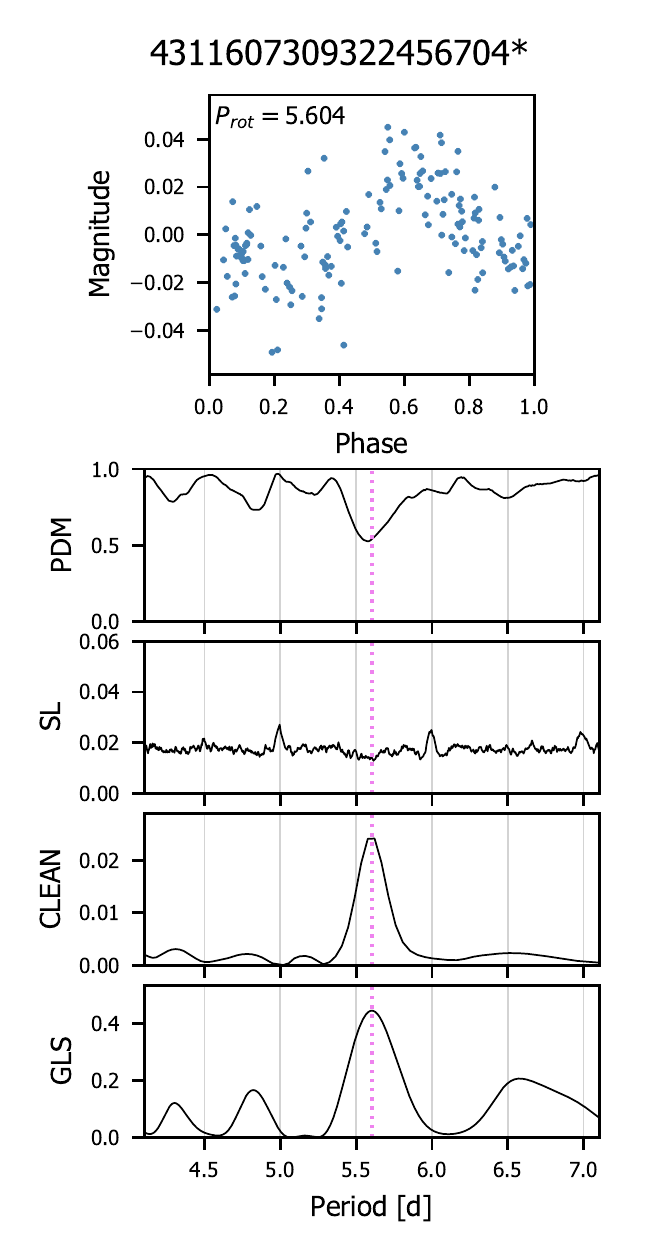}
     \includegraphics[width=0.3\linewidth]{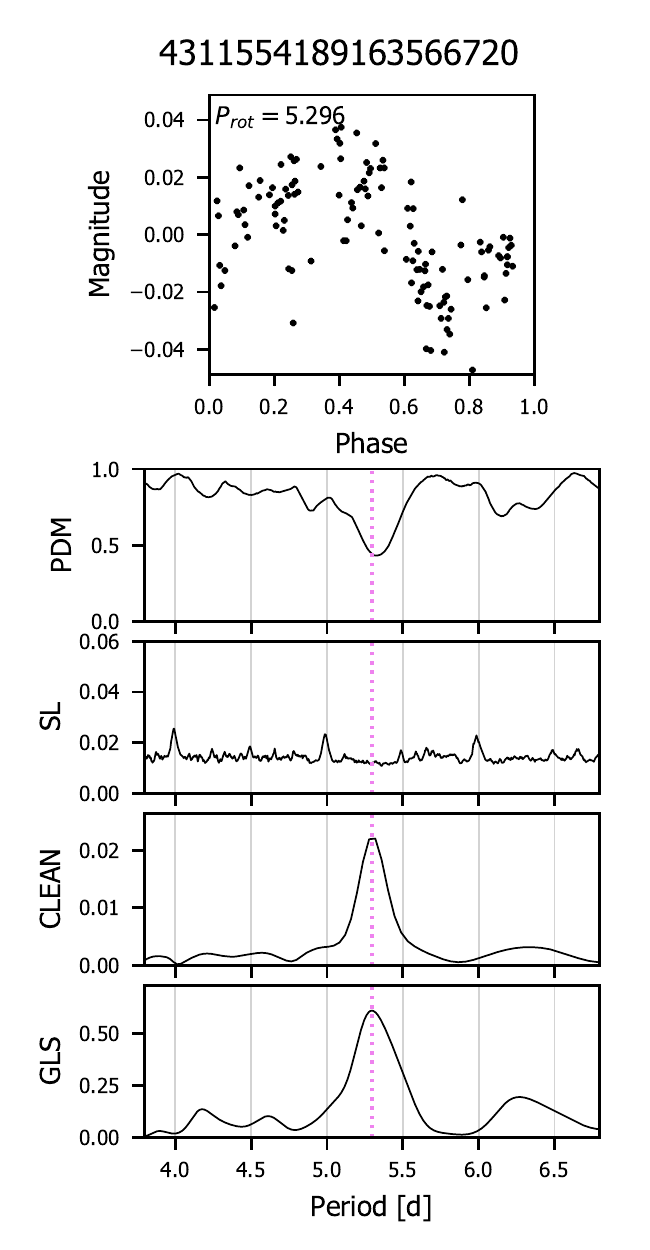}
     \includegraphics[width=0.3\linewidth]{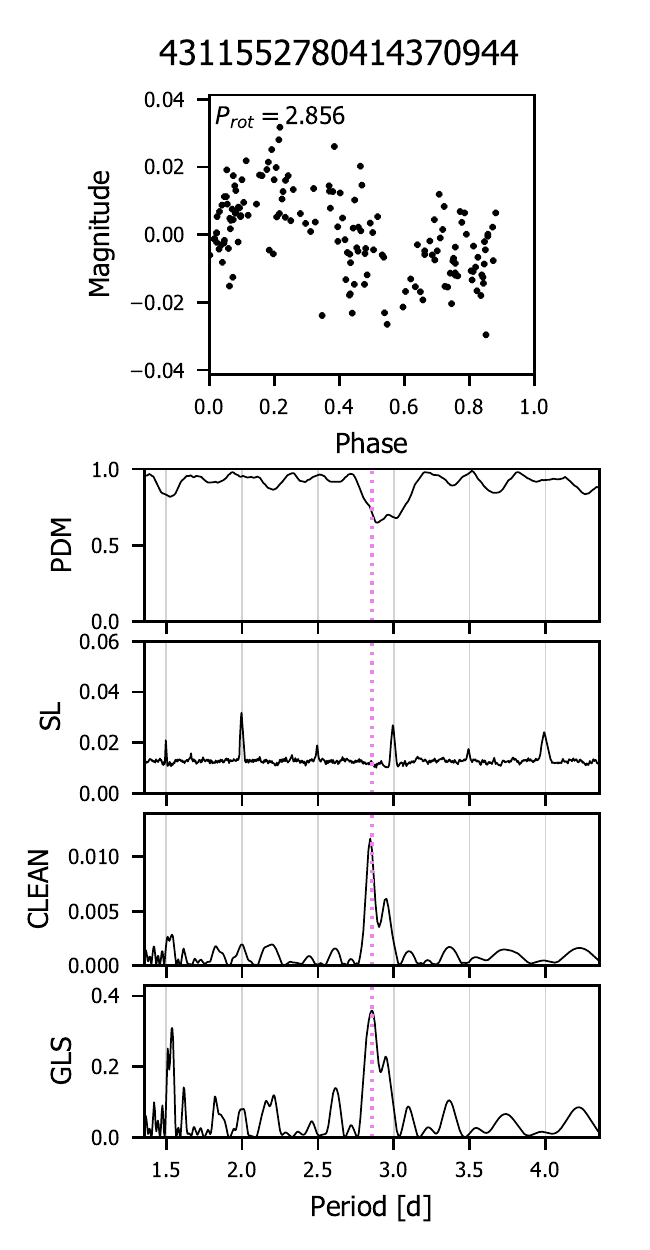}
     \caption{continued.}
     \label{fig:phased6}
\end{figure}

\begin{figure}\ContinuedFloat
     \centering
     \includegraphics[width=0.3\linewidth]{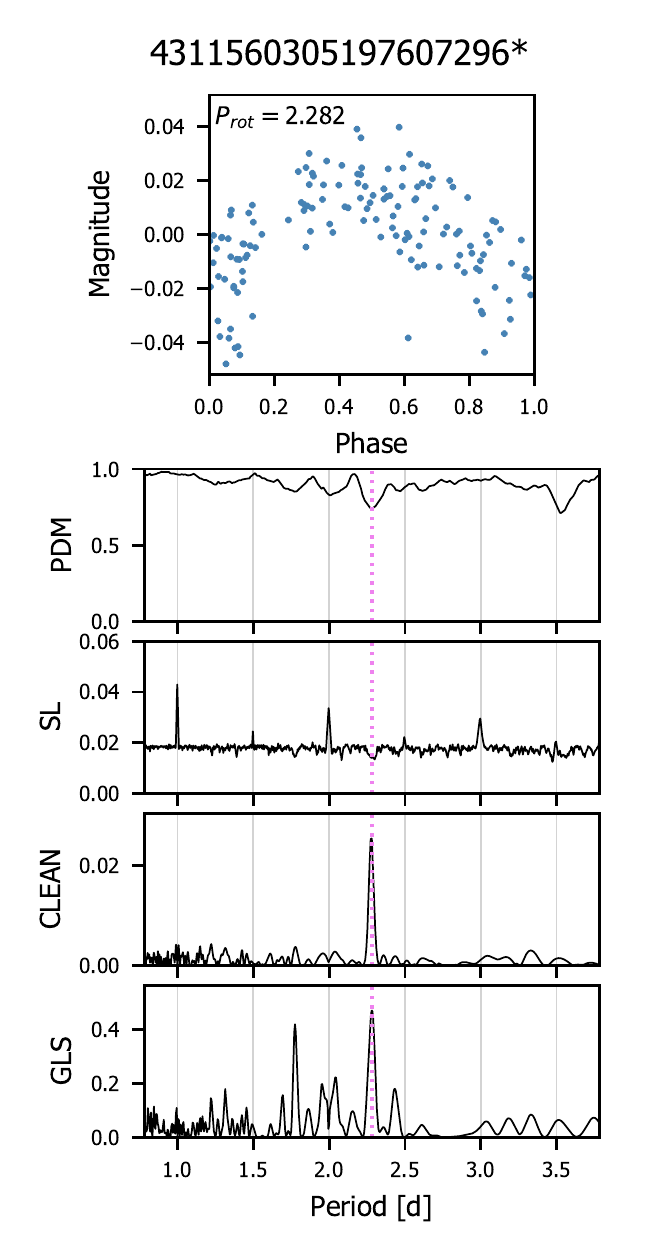}
     \includegraphics[width=0.3\linewidth]{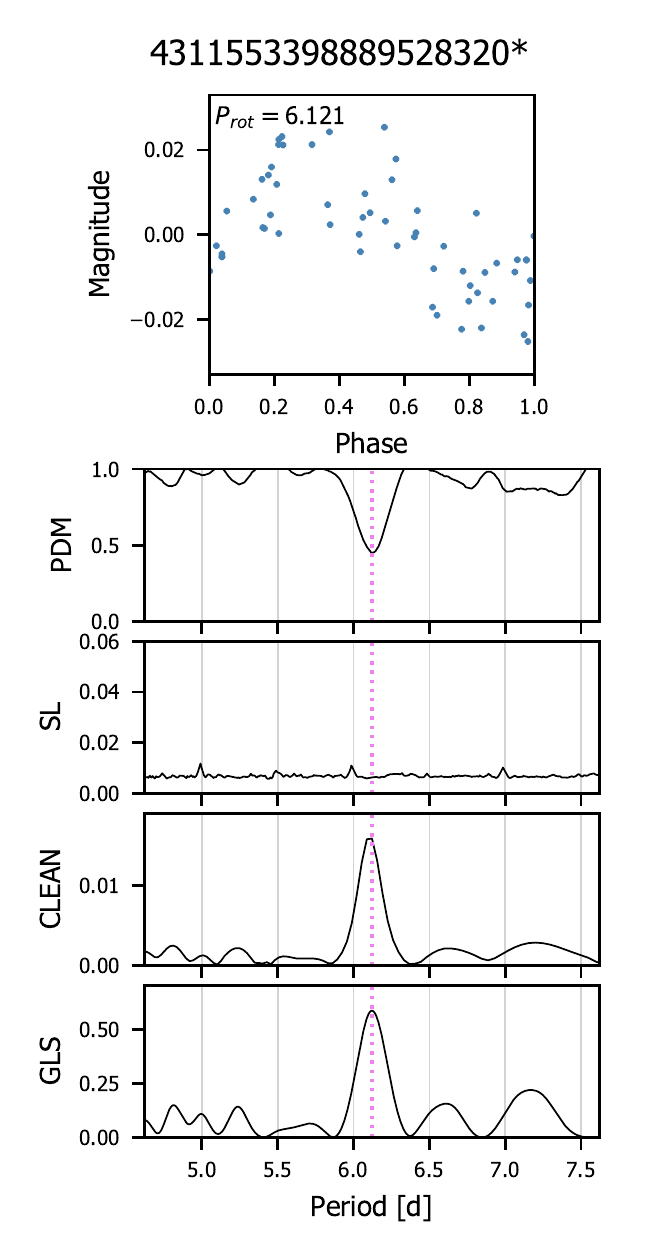}
     \includegraphics[width=0.3\linewidth]{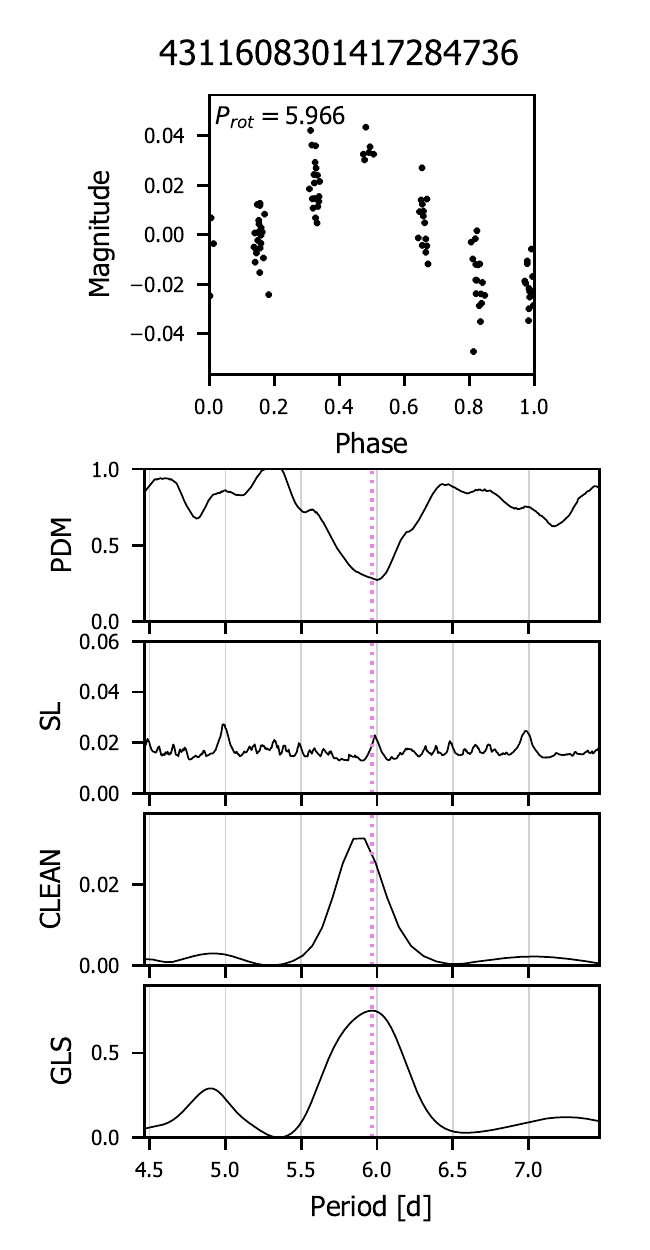}
     \includegraphics[width=0.3\linewidth]{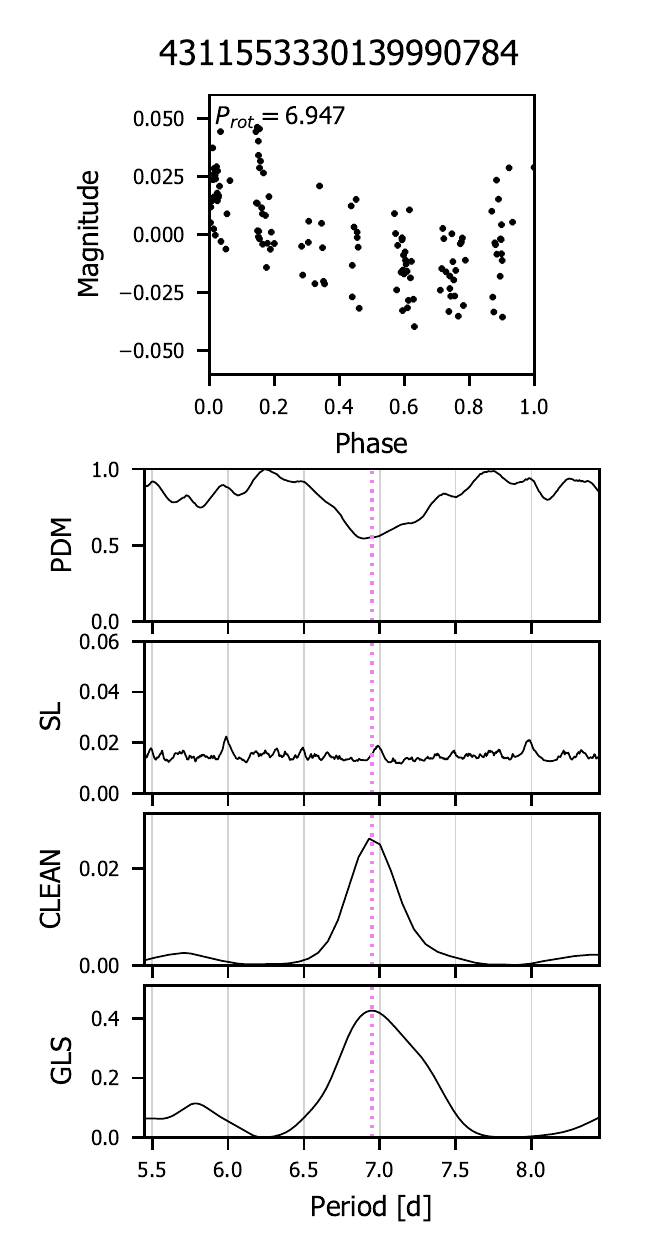}
     \includegraphics[width=0.3\linewidth]{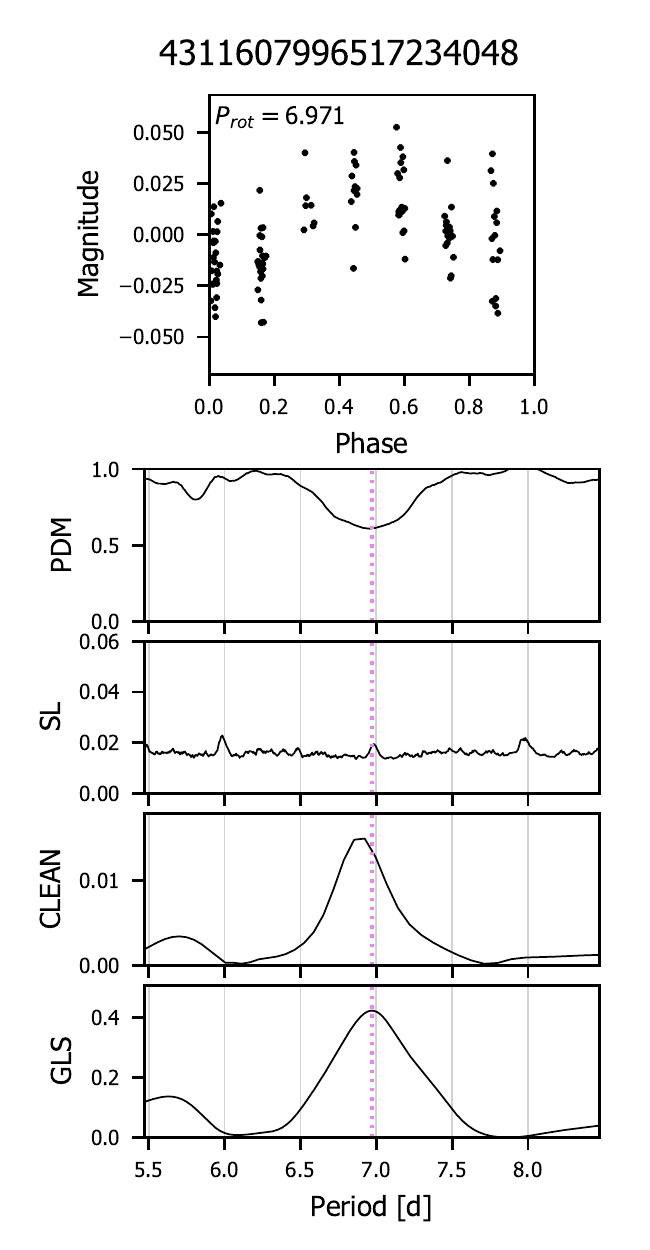}
     \includegraphics[width=0.3\linewidth]{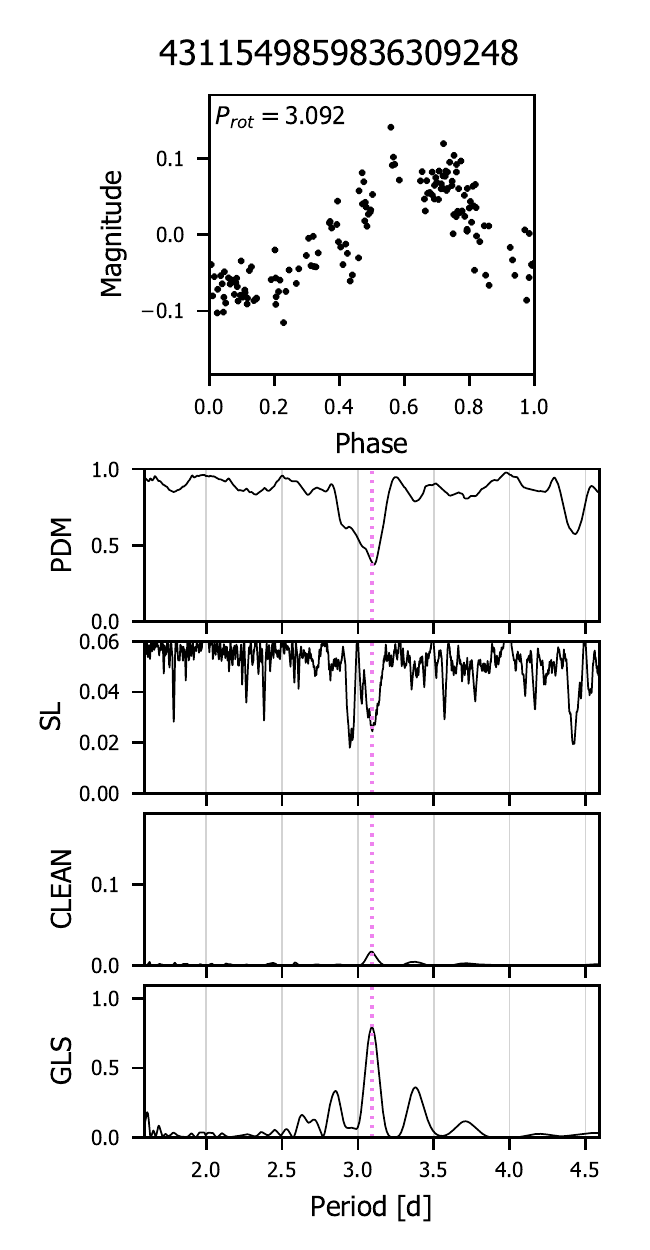}
     \caption{continued}
     \label{fig:phased7}
\end{figure}

\begin{figure}\ContinuedFloat
     \centering
     \includegraphics[width=0.3\linewidth]{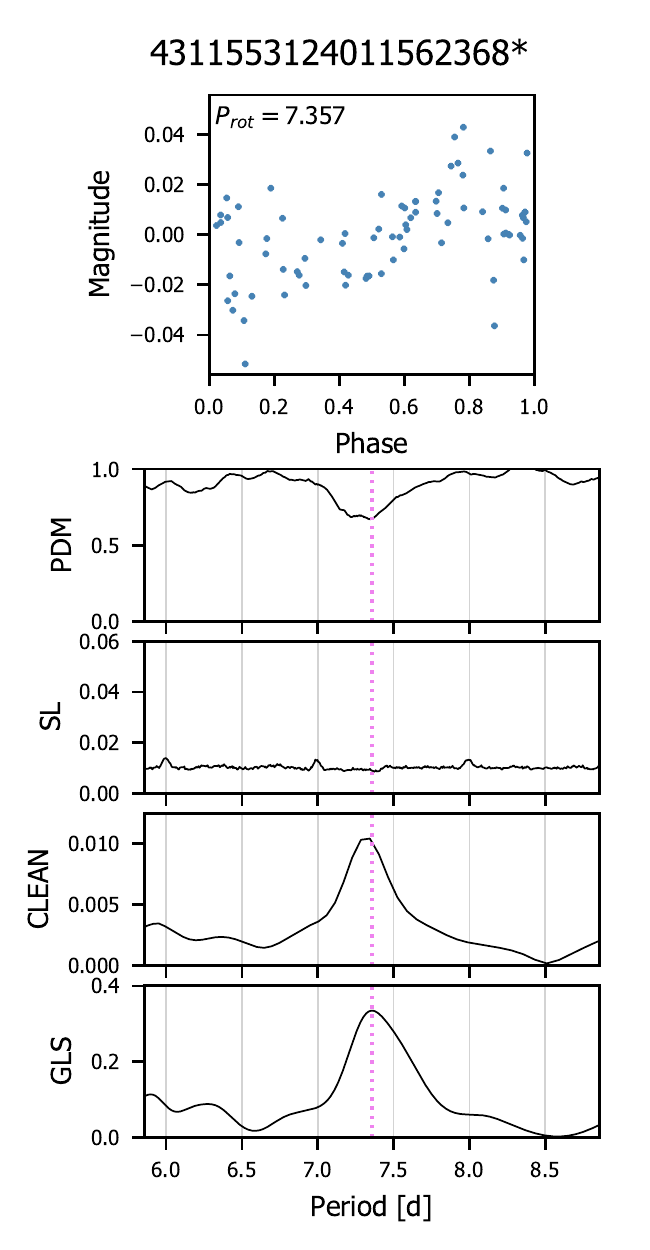}
     \includegraphics[width=0.3\linewidth]{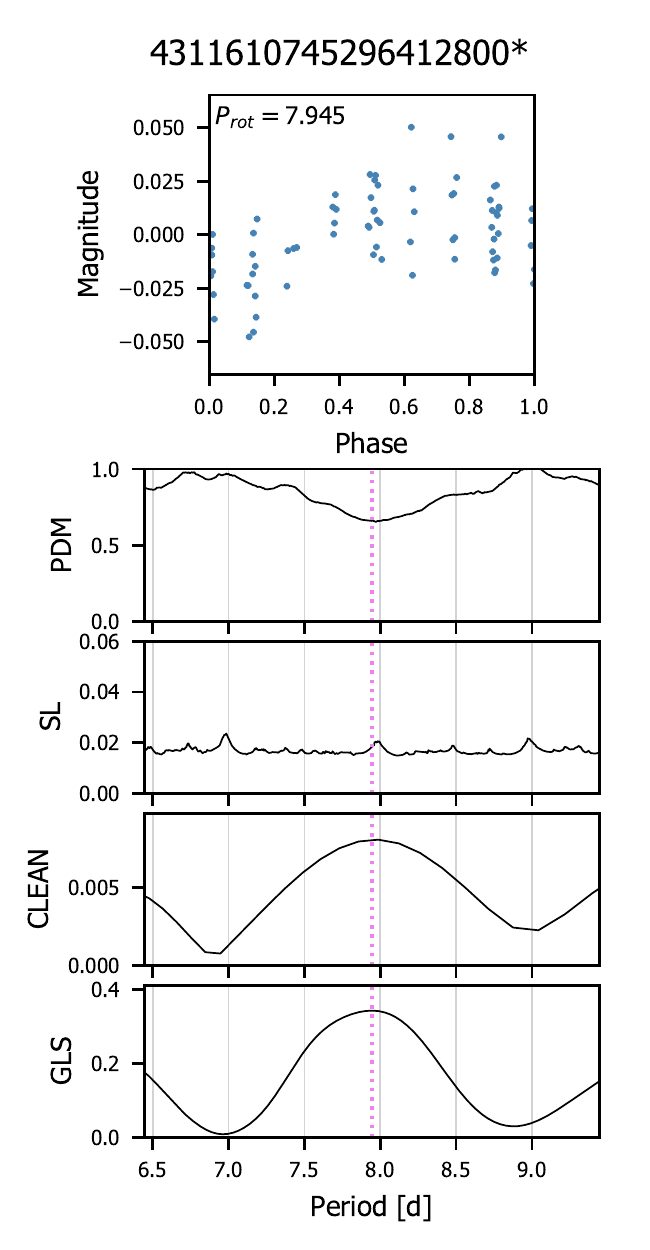}
     \includegraphics[width=0.3\linewidth]{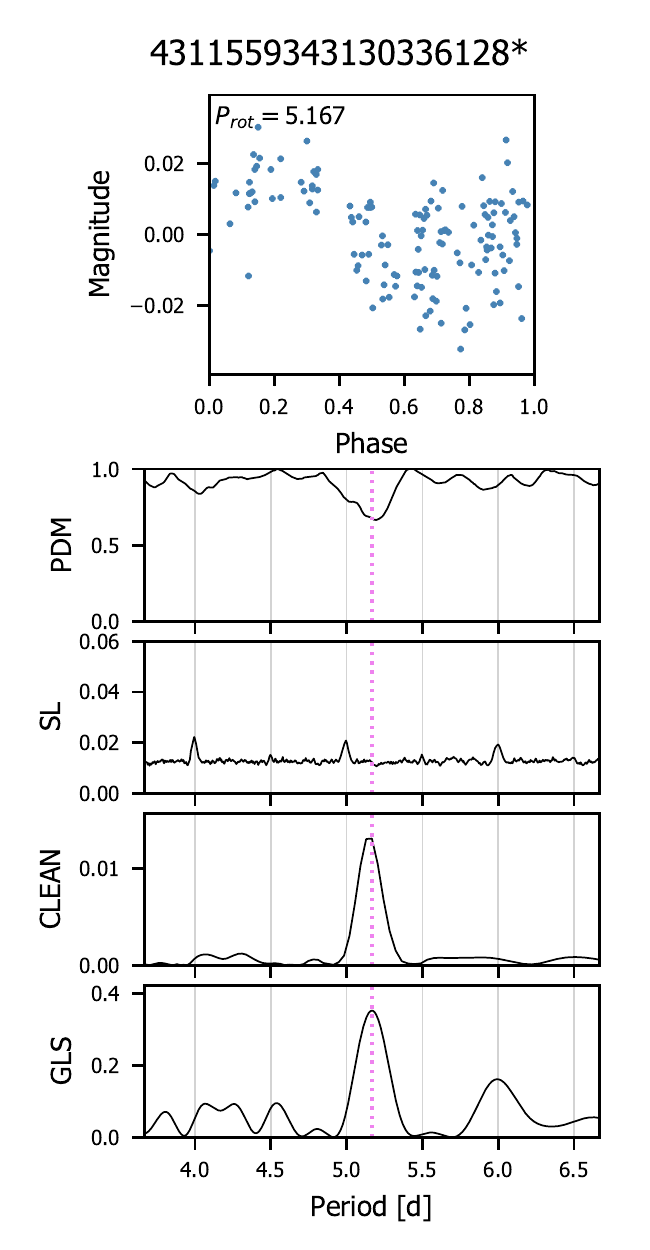}
     \includegraphics[width=0.3\linewidth]{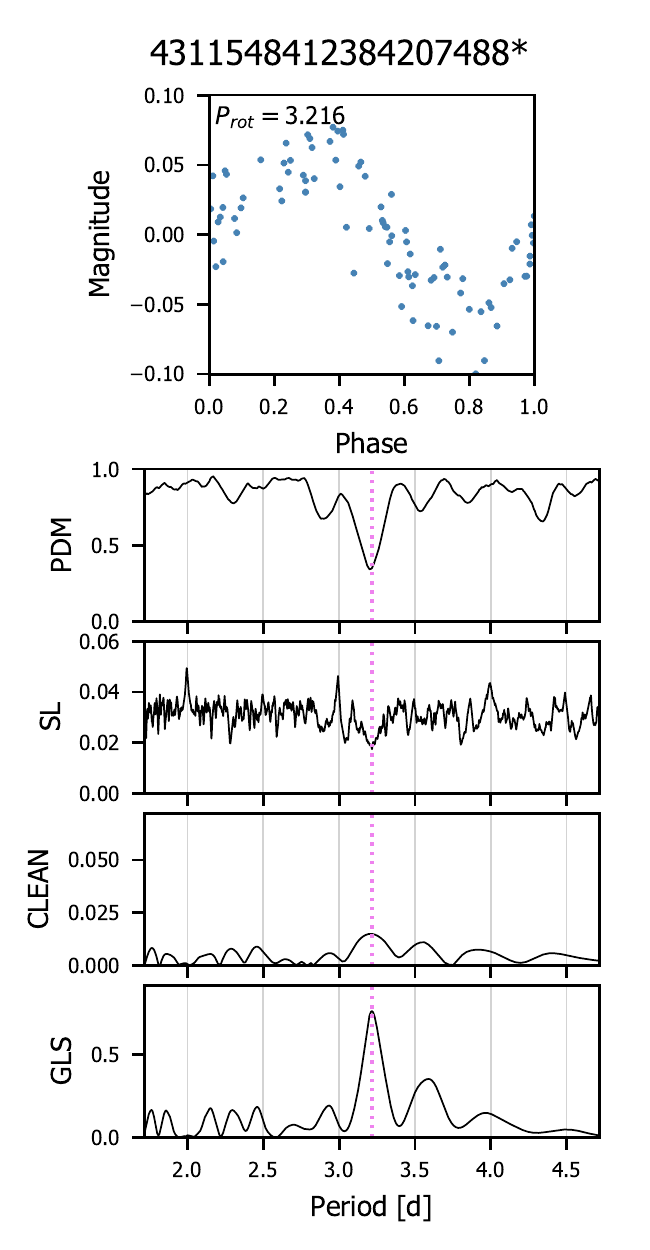}
     \includegraphics[width=0.3\linewidth]{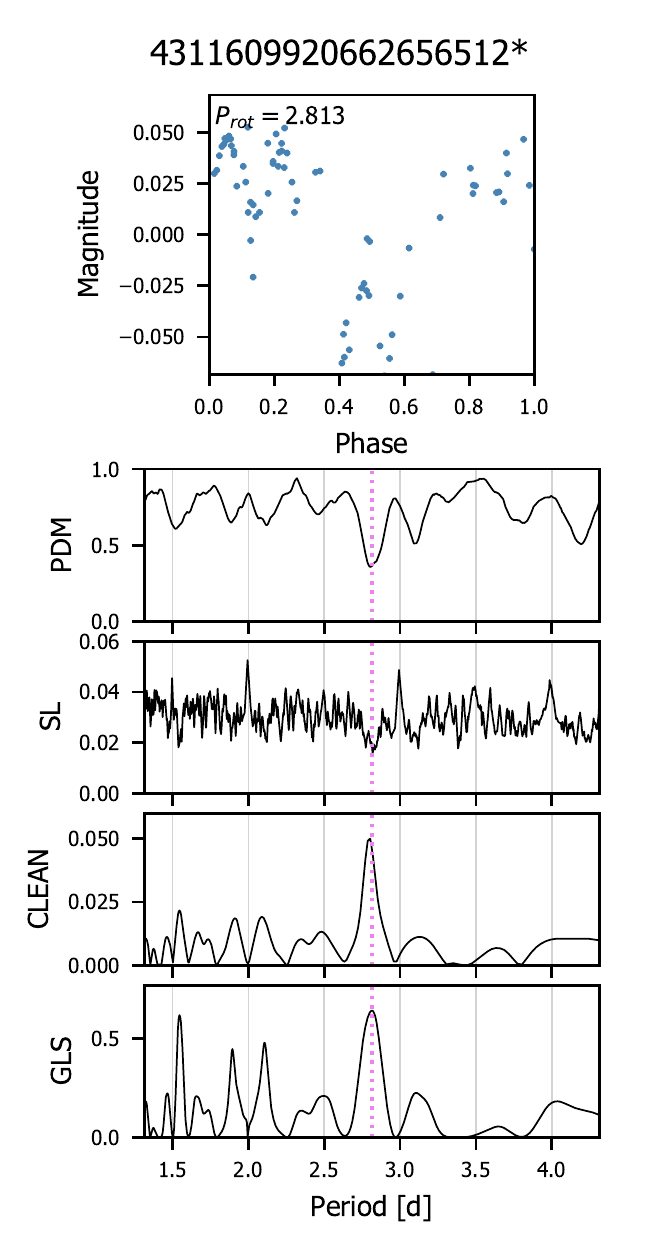}
     \includegraphics[width=0.3\linewidth]{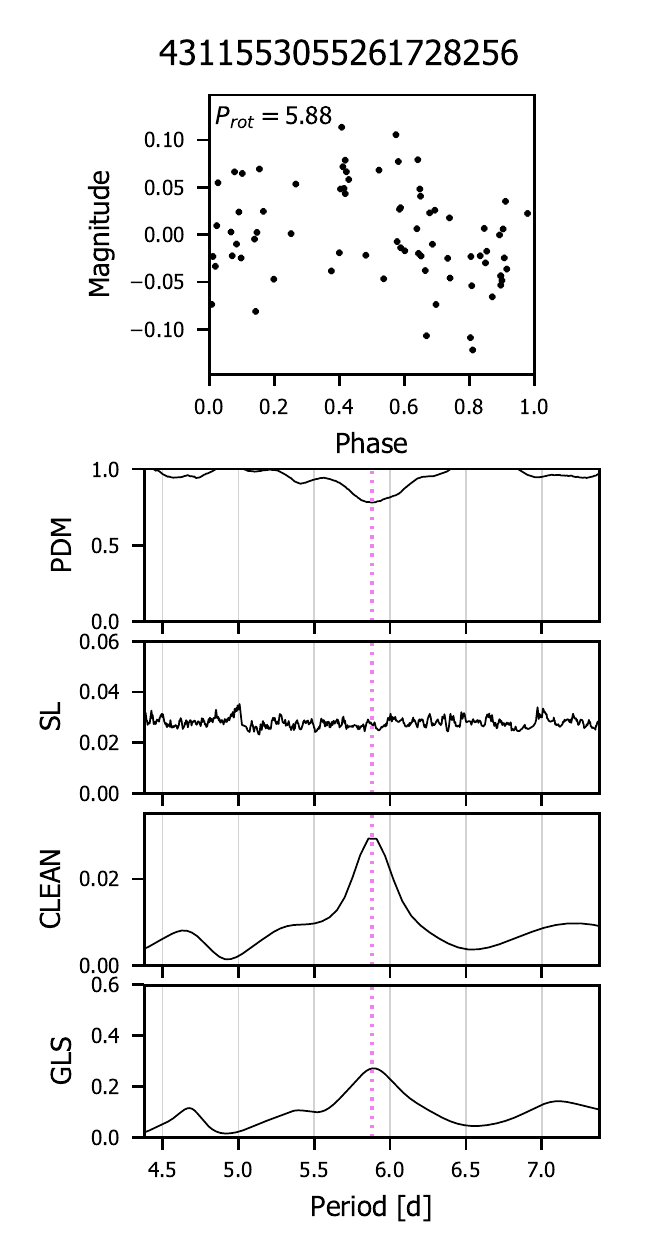}
     \caption{continued.}
     \label{fig:phased4}
\end{figure}

\begin{figure}\ContinuedFloat
     \centering
     \includegraphics[width=0.3\linewidth]{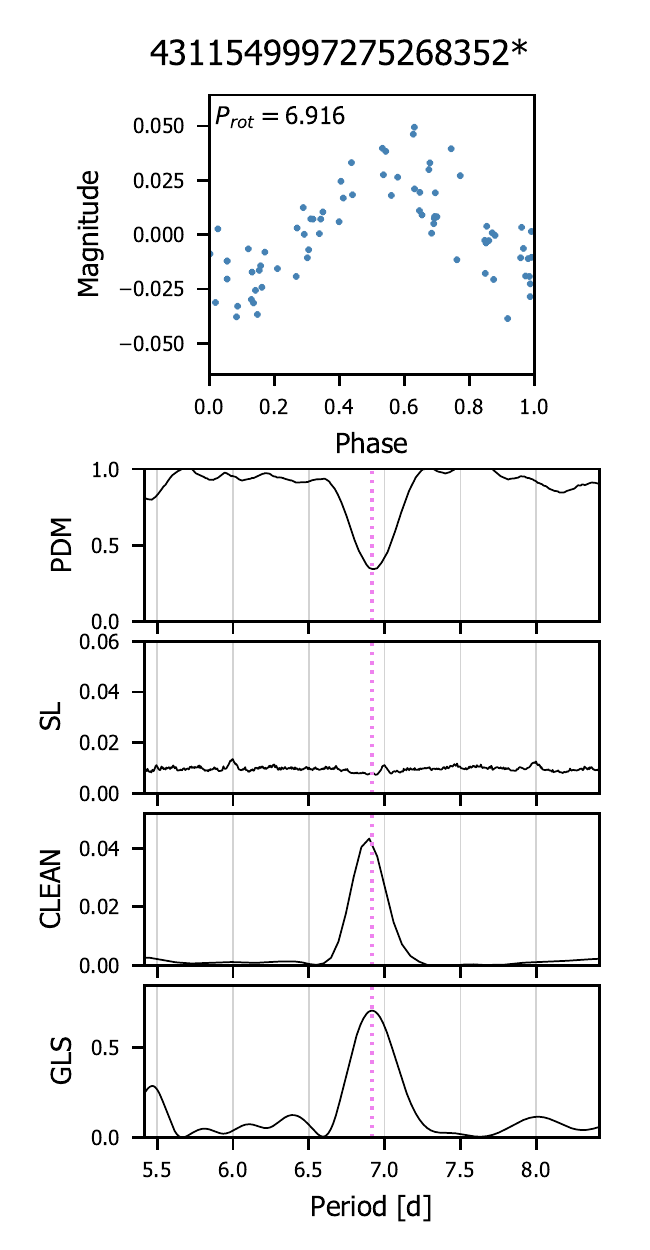}
     \includegraphics[width=0.3\linewidth]{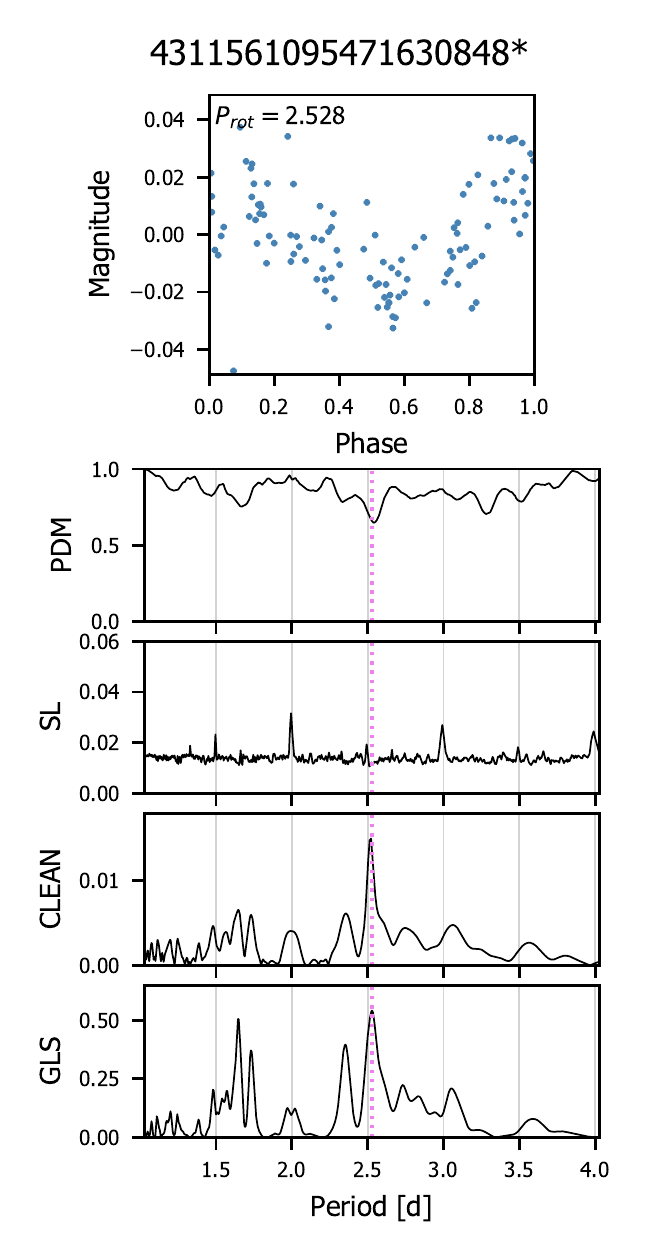}
     \includegraphics[width=0.3\linewidth]{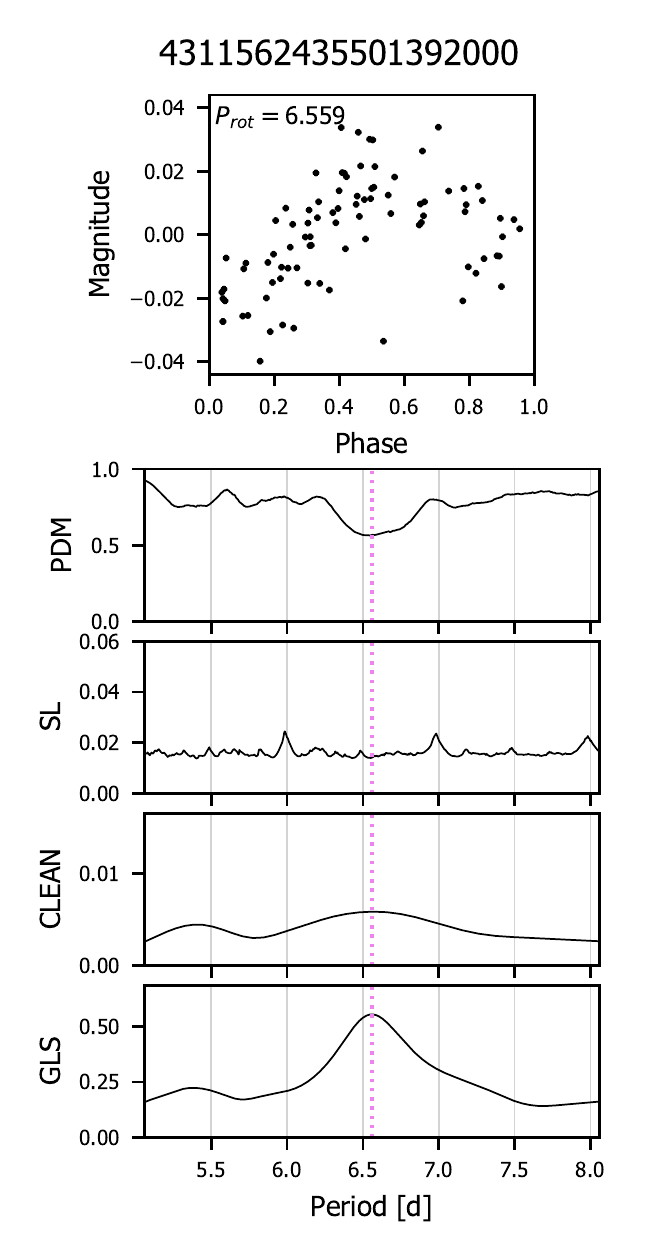}
     \caption{continued.}
     \label{fig:phased8}
\end{figure}
\end{appendix}
\end{document}